\documentclass[prb,twocolumn,aps,longbibliography]{revtex4-2}
\usepackage{graphicx,amsmath,xcolor}
\usepackage{hyperref}
\usepackage{subcaption}
\usepackage{float}
\usepackage{svg}
\usepackage{enumerate}

\usepackage{bibunits}
\usepackage{silence}
\usepackage{hyperref}
\usepackage{mathtools}
\usepackage[T1]{fontenc}
\DeclarePairedDelimiterX\innerp[2]{\langle}{\rangle}{#1\delimsize\vert\mathopen{}#2}%
\DeclarePairedDelimiterX\braket[2]{\langle}{\rangle}{#1\delimsize\vert\mathopen{}#2}%
\DeclarePairedDelimiterX\braketOP[3]{\langle}{\rangle}{#1\,\delimsize\vert\,\mathopen{}#2\,\delimsize\vert\,\mathopen{}#3}%
\DeclarePairedDelimiterX\ketbra[2]{\lvert}{\rvert}{#1\delimsize\rangle\!\delimsize\langle#2}%
\DeclarePairedDelimiterX\outerp[2]{\lvert}{\rvert}{#1\delimsize\rangle\!\delimsize\langle#2}%
\DeclarePairedDelimiterX\projector[1]{\lvert}{\rvert}{#1\delimsize\rangle\!\delimsize\langle#1}%
\newcommand{\comment}[1]{}
\newcommand{\subsecref}[1]{S-\Roman{section}.\arabic{subsection}}
\begin{document}
\defaultbibliography{mainbib}
\defaultbibliographystyle{apsrev4-2}
    \title{Mitigating disorder and optimizing topological indicators with vision-transformer-based neural
    networks in Majorana nanowires}
    \author{Jacob R. Taylor}
    \author{Sankar Das Sarma}
    \affiliation{Condensed Matter Theory Center and Joint Quantum Institute, Department of Physics, University of Maryland, College Park, Maryland 20742-4111 USA}
\begin{abstract}
Disorder remains a major obstacle to realizing topological Majorana zero modes (MZMs) in superconductor-semiconductor nanowires, and we show how deep learning can be used to recover topological MZMs mitigating disorder even when the pre-mitigation situation manifests no apparent topology. The disorder potential, as well as the scattering invariant ($T_V$) normally used to classify a device as topologically non-trivial are not directly measurable experimentally. Additionally, the conventional signatures of MZMs have proved insufficient due to their being accidentally replicated by disorder-induced trivial states. Recent advances in machine learning provide a novel method to solve these problems, allowing the underlying topology, suppressed by disorder, to be recovered using effective mitigation procedures.  In this work, we leverage a vision transformer neural network trained on conductance measurements along with a CMA-ES optimization framework to dynamically tune gate voltages mitigating disorder effects. Unlike prior efforts that relied on indirect cost functions, our method directly optimizes $T_V$ alongside additional local density of states-based topological indicators. Using a lightweight neural network variant, we demonstrate that even highly disordered nanowires initially lacking any topologically non-trivial regions can be transformed into robust topological devices.
\end{abstract}
\maketitle 
\begin{bibunit}
\textit{Introduction.\textemdash} 1D superconductor-semiconductor nanowires are widely regarded as a promising platform for topological quantum computation due to their potential to host Majorana zero modes (MZMs) \cite{lutchyn2010majorana,sau2010robustness,oreg2010helical,sau2010generic,das2023search,sarma2015majorana,lutchyn2018majorana}. These non-Abelian quasiparticles are protected by a "topological gap," exhibit resilience to local perturbations, and encode information through topologically non-local states, enabling robust gate operations via braiding \cite{kitaev2001unpaired}. However, a major and persistent challenge is the presence of disorder within these nanowires (caused mainly by charged impurities \cite{ahn2021estimating,woods2021charge}), which inhibits the realization and utilization of non-trivial topological states \cite{das2023search}.

Disorder presents a significant challenge for two major reasons. First, disorder, particularly when it exceeds the proximity-induced gap \cite{ahn2021estimating}, can destroy the non-trivial topology of MZMs \cite{das2023search}, which is necessary for protection and robust gate operations \cite{kitaev2003fault}. Second, disorder significantly confounds the identification of MZMs. Conventional approaches, such as zero-bias conductance peaks (ZBCPs) \cite{sengupta2001midgap} as a signature of MZMs fail, in the presence of strong disorder, where many properties thought to indicate MZMs are often produced by trivial disorder-induced Andreev bound states \cite{pan2020physical,liu2012zero,das2021disorder}. Many transport experiments have reported ZBCPs consistent with MZMs \cite{das2012zero,deng2012anomalous,mourik2012signatures,churchill2013superconductor,finck2013anomalous,nichele2017scaling,zhang2021large,yu2021non,song2022large,aghaee2023inas}, but it is generally agreed that most, if not all, observed ZBCPs result from trivial Andreev bound states \cite{das2023search}. Several experimental approaches have been proposed to address this issue, such as end-to-end correlated ZBCPs, bulk gap closing and reopening, and the topological gap protocol \cite{pikulin2021protocol,aghaee2023inas,pan2021three}, where a series of tests are conducted to declare a nanowire as topological. But, these are not guaranteed to be definitive as strong disorder often masks all signatures of topology \cite{das2023spectral,das2023density}.

The key problem is exemplified by the fact that even the theoretical gold standard for identifying MZMs, the scattering invariant $T_V$, which can uniquely determine the topology of an infinite wire \cite{akhmerov2011quantized,kitaev2001unpaired}, may be insufficient for assessing topology in a short disordered wire \cite{pan2024disordered,das2023spectral}. To make devices viable for fusion and braiding topological gate operations, it is necessary to supplement $T_V$ with additional operational indicators for disordered wires so as to ensure that accidental Andreev bound states in the bulk of the wire do not hinder topology. These proposed indicators rely on the local density of states (LDOS) to determine whether MZMs are localized at the wire's edges without any domain walls in the bulk. Since these conditions must be independently verified to ensure the presence of true topological MZMs, it is necessary to assess devices not only using $T_V$ but these additional operative LDOS indicators, such as $I_2$ and $I_1$ discussed in \cite{pan2024disordered,taylor2025vision}. Recent work has explored the use of machine learning to characterize \cite{taylor2025vision,taylor2024machine,cheng2024machine}, correct \cite{thamm2023machine,thamm2024conductance}, and diagnose \cite{taylor2025vision,cheng2024machine} disorder in Majorana nanowires. 

The important question is whether disorder can be suppressed and corrected elucidating the underlying pristine topology using gate operations since materials fabrication of  disorder-free wires has turned out to be problematic. We show in the current work how topology can be restored by using local voltage gates to adjust the potential and compensate for disorder \cite{thamm2023machine,thamm2024conductance}. The main challenge is that disorder in a nanowire device cannot be measured directly. While neural networks theoretically allow direct determination \cite{taylor2024machine}, scaling to the necessary size remains a challenge. Despite this hurdle, indirect methods have been explored, including using CMA-ES \cite{hansen2016cma} a genetic algorithm-like optimization procedure, to address disorder effects. These approaches rely on alternative indicators, such as coherent transmission \cite{thamm2023machine} or a simplified topological gap protocol (TGP) \cite{thamm2024conductance}, to define an effective cost function. When optimized, this function is expected to increase the probability of realizing true MZMs. However, the relationship between these topological indicators and $T_V$, a fundamental requirement for MZMs, remains poorly understood or insufficient. 

Given the challenges of indirect cost functions and, more importantly, the classification of these devices, interest has grown in using advanced neural network methods to determine $T_V$ \cite{taylor2025vision,cheng2024machine}. Recent results demonstrate one can infer whether a device hosts a topologically non-trivial phase using only conductance measurements. One study goes even further, suggesting that devices can, in theory, be topologically classified with arbitrarily high confidence, providing not only the full phase diagram of $T_V$ but also the LDOS based Majorana indicators \cite{taylor2025vision}. 
In this work, we utilize the results of \cite{taylor2025vision} to predict topological indicators with high accuracy from conductance measurements and then perform a CMA-ES mitigation procedure to clearly bring out the underlying topology which is hidden by the disorder before mitigation.  We create topology from nothing using machine learning! We consider a wire with a series of adjustable gate voltages and optimize directly for $T_V$, combined with additional LDOS-based operational indicators. 
We demonstrate, through simulations, that even for disorder levels far exceeding those expected in experiments and correlation lengths worse than anticipated, our scheme can take wires with no topologically non-trivial regions and mitigate their disorder to create robust regions of topological non-triviality. Our scheme indeed produces topology from nothing in realistic disordered wires.

\textit{Physical Model.\textemdash }
We model the 1D semiconductor Majorana nanowire using a Bogoliubov–de Gennes Hamiltonian \cite{lutchyn2010majorana}:
\begin{multline}\label{eq:1}
H=\left(-\frac{\hbar^2}{2m^*}\partial_x^2-i\alpha\partial_x\sigma_y-\mu+V_{\text{dis}}(x)+V_g(x)\right)\tau_z+\\\frac{1}{2}g\mu_BB\left(\sigma_x\cos\theta+\sigma_y\sin\theta\right)+\Sigma(\omega)\end{multline} 
where $\Sigma(\omega)=-\gamma \frac{\omega+\Delta_0\tau_x}{\sqrt{\Delta_0^2-\omega^2}}$ is the self-energy generated by integrating out the superconductor proximity-coupled to the semiconductor~\cite{sau2010robustness}. The matrices $\sigma_{x,y,z}$ and $\tau_{x,y,z}$ represent the spin and superconducting particle-hole degrees of freedom, respectively. The above Hamiltonian is written in a basis where the Bogoliubov quasiparticles of the superconductor are described by a Nambu spinor with the structure $\psi(x)=(u_{\uparrow}(x), u_{\downarrow}(x),-v_{\downarrow}(x),v_{\uparrow}(x))^T$, where $\uparrow,\downarrow$ denote spin, and $u,v$ represent the particle-hole components of the quasiparticle. The frequency $\omega$ in the self-energy corresponds to the energy (in units where $\hbar=1$) of the Bogoliubov quasiparticle under consideration. Specific parameter values match those in \cite{taylor2025vision}(e.g., effective mass $m^*=0.03 m_e$, superconducting pairing potential $\gamma=0.15,\text{meV}$, g-factor $g=25$, superconducting gap $\Delta_0=0.12,\text{meV}$, temperature $T=50,\text{mK}$, and device length $L=3,\mu\text{m}$) chosen to fit both non-superconducting transport characterization and superconducting transport properties \cite{pan2021three,woods2021charge,das2023spectral}. The remaining parameters, Rashba spin-orbit coupling $\alpha$, disorder $\delta_{\text{dis}}(x)$ ($\delta_{dis}$ for magnitude), and its correlation length $L_{\text{dis}}$ are treated as unknowns. The wire is assumed to be in proximity to a series of gates (which mitigate disorder as prescribed by machine learning) with varying voltages, resulting in a modification to the potential $V_g(x)$. The gates are approximated as having no gaps between them and being separated from the wire by $z_0=87$ nm, resulting in a broadening effect and the following potential: $$\begin{aligned}
V_g(x)=\mathcal{F}^{-1}\left[\mathrm{e}^{-|q|z_0}\mathcal{F}\left[\sum_{j=1}^{N_\mathrm{g}}V_j\chi_j(y)\right]\right]
\end{aligned}$$

$$\begin{aligned}V_{j} & =\frac{b_0}{2}+\sum_{k=1}^{\left\lfloor\frac{N_\mathrm{g}-1}{2}\right\rfloor}a_k\sin\left(\frac{2\pi}{N_\mathrm{g}}kj\right)
 & +\sum_{k=1}^{\left\lfloor\frac{N_\mathrm{g}}{2}\right\rfloor}b_k\cos\left(\frac{2\pi}{N_\mathrm{g}}kj\right).
\end{aligned}$$

where $\mathcal{F}$ is the Fourier transform, $N_g=20$ and $\chi_j(y)$ is 1 if $y$ corresponds to a position within the $j$th gate's region and 0 otherwise. Characterizing $V_j$ by Fourier components $\vec{\beta}=[\vec{a},\vec{b}]$ was found more effective in previous work \cite{thamm2023machine}. 

To determine whether a wire possesses a topological phase, we use three indicators: the scattering matrix invariant $T_V = \det(S)$ and two LDOS ($\rho_{LDOS}$) based indicators, $I_2$ and $I_1$, originally introduced in \cite{pan2024disordered} and then used in a modified form in\cite{taylor2025vision}. A device passes the topological test if all three indicators are negative simultaneously. Thus, we define the combined indicator $I_C$ as follows:
\begin{equation}
I_C= -(\theta(-I_1)\theta(-I_2)\theta(-T_V))^{1/3} 
\end{equation} where $\theta(x)$ is the Heaviside function. These indicators are only valid for the ($\mu$,$B$) points at which they are computed. To form a single-valued cost function, we take their average: $C_{T_V}(V_g)=\sum_i^{N_{points}} \frac{T_V(\mu_i,B_i)}{N_{points}}$ and $C_{C}(V_g)=\sum_i^{N_{points}} \frac{I_C(\mu_i,B_i)}{N_{points}}$ where $N_{points}$ is the number of ($\mu_i$,$B_i$) measurement configurations. We define $\tilde{C}_{C}(V_g) = C_C + \gamma C_{T_V}$, where $\gamma=1/100$ lets optimization initially rely on $C_{T_V}$ until points with $I_C \neq 0$ appear. Since many devices have $I_C = 0$ over the entire region, $T_V$ must be tuned first to reach a regime where $I_C$ becomes relevant and dominates. We emphasize that $I_C$ is an extremely stringent indicator, being a product of three independent indicators for MZMs.

\textit{Neural Network Model and Training Data.\textemdash }
To determine our cost function and, consequently, the topological indicators, a neural network almost identical to that presented in \cite{taylor2025vision} is used. The network operates by taking a series of conductance measurements of a wire under different experimental configurations to predict the topological indicators. Three configuration parameters are varied: the chemical potential $\mu$ and the magnetic field $B$. For each ($\mu,B,V_{Bias}$) setup, four conductance measurements are obtained: [$G_{RR}, G_{LR}, G_{RL}, G_{LL}$]. The network makes use of a "lite" version of the previous vision transformer network \cite{taylor2025vision} by reducing the number of measurements and directly predicting $C_{T_V}$ and $C_{C}$. Additional details are shown in supplementary section \ref{sec:neuralnetworkdetails} {(see also references \cite{dosovitskiy2020image,taylor2024neural,fcmaes2021} therein).}  

\textit{The Algorithm.\textemdash }
The mitigation process closely follows that in \cite{thamm2023machine,thamm2024conductance}, with CMA-ES optimizing the gate voltages. The algorithm searches for a gate configuration characterized by $\vec{\beta}$ to minimize $C^{pred}_{T_V}$, iteratively refining $\vec{\beta}$ by requesting evaluations of $C^{pred}_{T_V}$ for different gate setups. Each evaluation involves setting the nanowire's gate voltages and performing conductance measurements as input to the prediction neural network, which estimates the cost function $C^{pred}_{T_V}$. CMA-ES then updates $\vec{\beta}$, and the process is repeated, gradually optimizing $C^{pred}_{T_V}$. The bottleneck in this optimization process is device simulations with KWANT\cite{groth2014kwant}, as each CMA-ES generation must run sequentially. This constraint limits the resolution of device measurements available for the neural network. Experimentally, this limitation does not exist, as conductance measurements are performed in scans, routinely producing high-resolution data at little additional cost, allowing the full network from \cite{taylor2025vision} to be used and likely improving prediction accuracy. For details on CMA-ES usage and parallelization see section S-II.C. 

    \begin{figure}[]
     \centering
     \begin{subfigure}[b]{0.9\columnwidth}
         \centering
         \includegraphics[width=\textwidth]{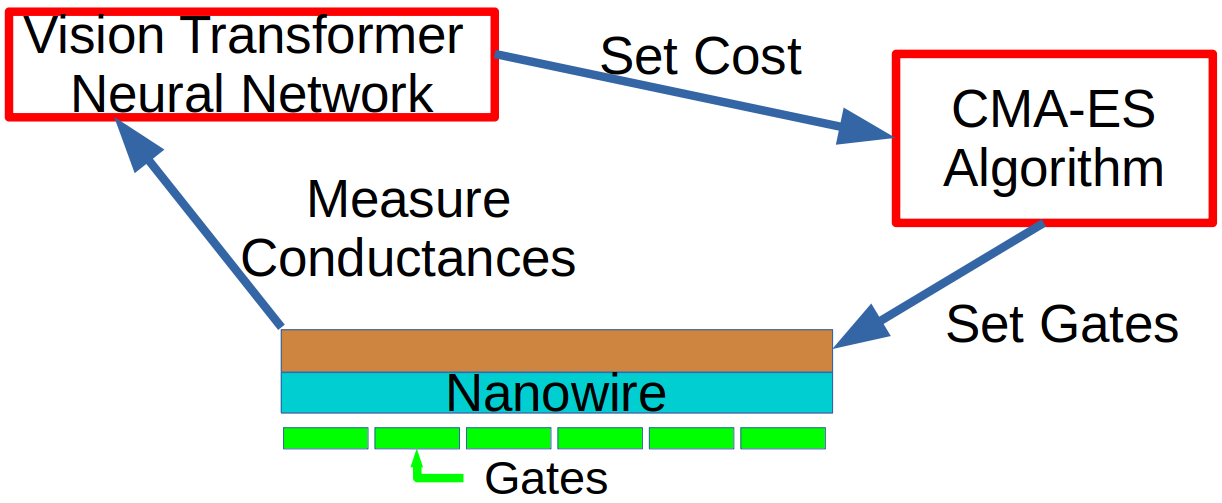}
         \caption{}
     \end{subfigure}
     \caption{Diagram of disorder mitigation process. The process consists of measuring an array of conductances on a wire with set gate voltages (initially random), these conductances are then fed into the vision transformer network. The neural network produces a cost function based on predicting $C^{pred}_{T_V}$ or $C^{pred}_{C}$ feeding the result into CMA-ES. CMA-ES provides a new set of gate voltages and the cycle repeats allowing mitigation of the effects of disorder within the wire. }
     \label{fig:cycle}
\end{figure}

\textit{Results.\textemdash } We first evaluate our mitigation method on a wire with $\delta_{dis} = 2.5$ meV, significantly exceeding the expected experimental upper limit of $1.5$ meV in \cite{aghaee2023inas}. $L_{dis}$ is set to $70$ nm, their minimum estimated value. Initially, the wire has no passing regions in $I_C$ and only a few isolated points passing in $T_V$. After 2000 iterations of the mitigation cycle (see Fig. \ref{fig:2,5BAIndicators}), using $C^{pred}_{TV}$ as the cost function, $C_{TV}$ improves from $0.961 \rightarrow 0.388$. The effectiveness is evaluated with the actual $C_{TV}$ rather than $C_{TV}^{pred}$. More impressively, $C_C$ improves from $0.000 \rightarrow -0.077$, meaning the device transitions from having no passing points in $I_C$ (and no useful non-trivial topology) to large passing segments. The mitigation scheme successfully transforms a wire initially too disordered to be at all useful into one that is robustly topological. Post-mitigation results (Fig. \ref{fig:2,5BAIndicators}) show a large region passing $I_C$.

\begin{figure}[]
     \centering
     \begin{subfigure}[b]{0.99\columnwidth}
         \centering
         \includegraphics[width=\textwidth]{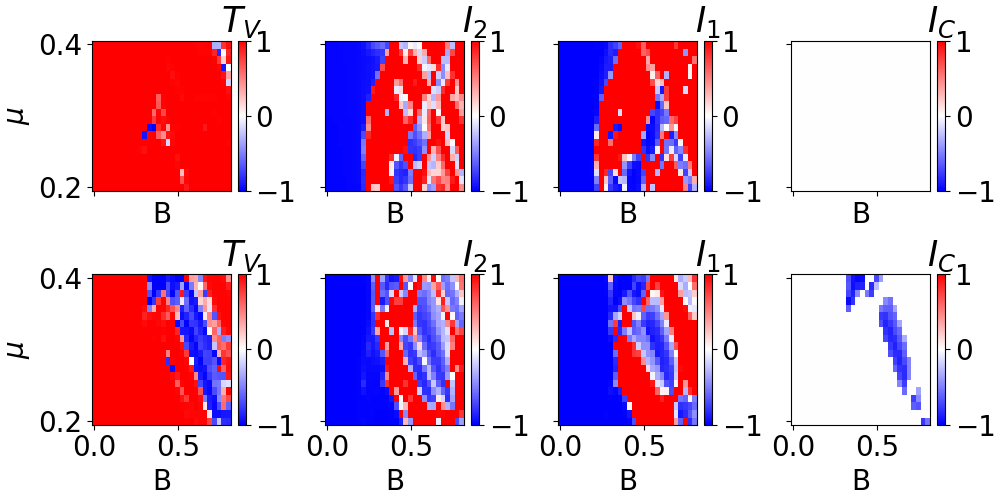}
         \caption{}
     \end{subfigure}
     \caption{Majorana indicators before and after disorder mitigation for a representative wire realization with $\delta_{dis}=2.5 meV$ and $L_{dis}=70nm$. The first (second) row shows $T_V$, $I_2$, $I_1$ and $I_C$ in that order before (after) mitigation for different $\mu$ and $B$. The mitigation scheme was ran for 2000 iterations using only $T_V$.}
     \label{fig:2,5BAIndicators}
\end{figure}

\begin{figure}[]
     \centering
     \begin{subfigure}[b]{0.99\columnwidth}
         \centering
         \includegraphics[width=\textwidth]{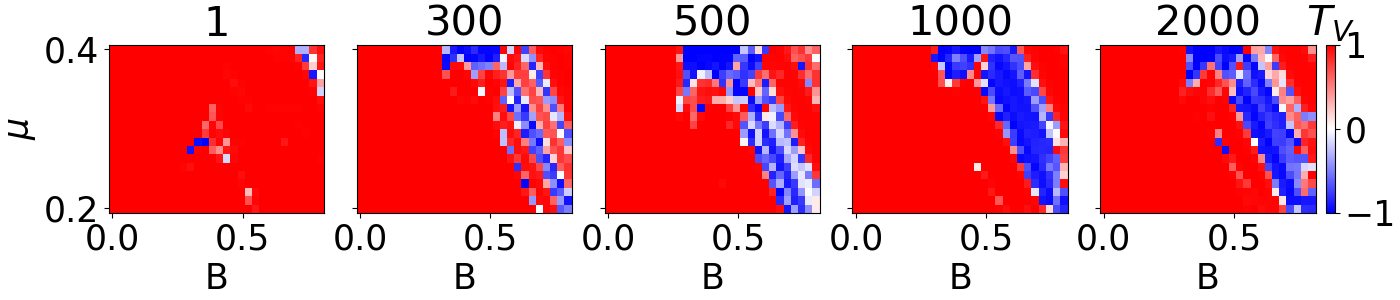}
         \caption{}
     \end{subfigure}
          \begin{subfigure}[b]{0.99\columnwidth}
         \centering
         \includegraphics[width=\textwidth]{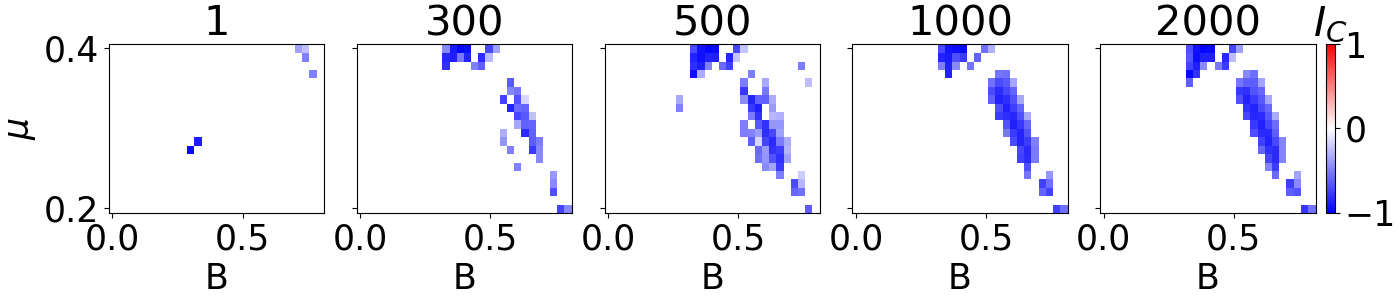}
         \caption{}
     \end{subfigure}
    \begin{subfigure}[b]{0.49\columnwidth}
         \centering
         \includegraphics[width=\textwidth]{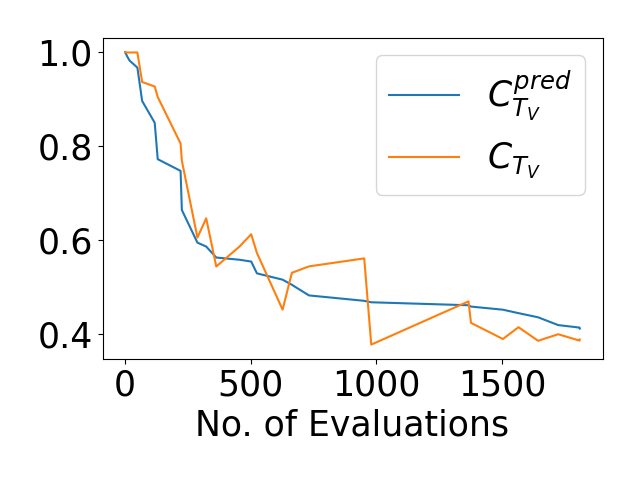}
         \caption{}
     \end{subfigure}
         \begin{subfigure}[b]{0.49\columnwidth}
         \centering
         \includegraphics[width=\textwidth]{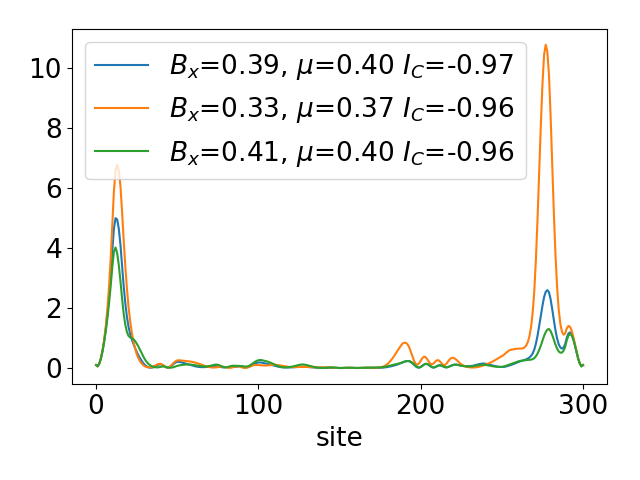}
         \caption{}
     \end{subfigure}
     \caption{(a) $T_V$ and (b) $I_C$ for different values $\mu$ and B with 1 (or 1 incomplete generation), 300, 500, 1000 and 2000  mitigation iterations (where one iteration consists of conductance measurements for a single gate voltage array) for a wire with $\delta_{dis}=2.5 meV$ and $L_{dis}=70nm$.  (c) Shows the neural network predicted cost $C^{pred}_{T_V}$ and actual cost $C_{T_V}$ for increasing numbers of iteration steps. (d) Shows the $\rho_{LDOS}(x)$ across the wire at a few selected points post mitigation.
     }
     \label{fig:timeresults}
\end{figure}

Fig. \ref{fig:timeresults} shows how device performance improves with mitigation iterations. Initial improvements are rapid, followed by gradual convergence in both cost functions. In this case, $C^{pred}_{T_V}$ and $C_{T_V}$ converge closely, though this is not generally true. Even in Fig. \ref{fig:timeresults}, $C^{pred}_{T_V}$ and $C_{T_V}$ align more closely in early iterations, when $\delta_{dis}(x) + V_g(x)$ resembles the training data  (i.e mostly random or zero $V_g$). As the mitigation scheme tunes gate voltages, differences can occur due to neural network hallucinations. Typically, both cost functions converge with some deviation between them. Neural network hallucinations can sometimes occur, where continued optimization eventually slightly worsens outcomes. However, this happens after significantly improving the $I_C$ and $T_V$ phases. CMA-ES may act adversarially, making minimal changes to actual output while significantly altering predicted output. This can be mitigated by limiting the number of iterations (early stopping) or restarting/re-initializing $\vec{\beta}$. While, the likelihood of hallucination increases at higher disorder levels, our method already succeeds at disorder levels beyond those expected in experiments, making this a minor concern. In Fig. \ref{fig:timeresults}, improvement stops after 1000 evaluations, with significant gains by 500.

\begin{figure}[]
     \centering
     \begin{subfigure}[b]{0.99\columnwidth}
         \centering
         \includegraphics[width=\textwidth]{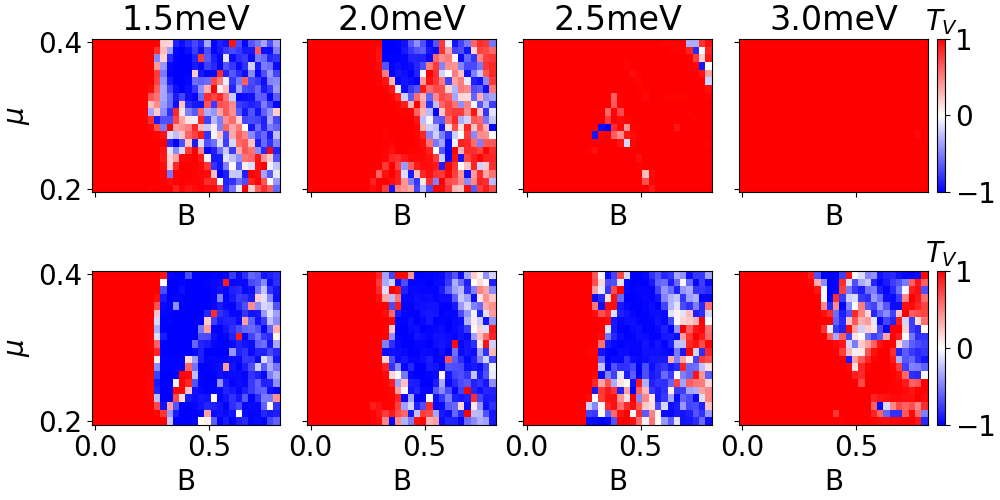}
         \caption{}
     \end{subfigure}
     \caption{$T_V$ plotted against $\mu$ and B for representative wire realizations of different disorder magnitudes and fixed $L_{dis}=70$nm. The first and second rows consists of $T_V$ before and after the mitigation scheme respectively.}
     \label{fig:MagComp}
\end{figure}

We consider wire realizations with varying $\delta_{dis}$ and $L_{dis}$. For $L_{dis} = 70$ nm, Fig. \ref{fig:MagComp} shows that disorder can be mitigated effectively up to $3.0$ meV. Beyond this, the current setup fails around $4.0$ meV, likely due to the fixed $N_g=20$ and bounds on $\vec{\beta}$ magnitudes. Fig \ref{fig:MagComp} shows different $T_V$ diagrams across mitigation runs can arise even with the same device, making success at $(\mu,B)$ points unpredictable. 
Success in $T_V$ does not imply success in $I_C$. At 3.0 meV in Fig. \ref{fig:MagComp}, $T_V$ mitigation improves $C_{T_V}$ from $1.00 \rightarrow 0.499$, yet $C_C$ remains at $0.0$. 
In general, optimizing $C_{T_V}$ does not guarantee regions with negative $I_C$, though it is often the case. Better results can be achieved either by starting with $\tilde{C}_C$ or continuing optimizing $\tilde{C}_C$. Using $\tilde{C}_C$ often significantly improves mitigation when $C_{T_V}$ alone fails to produce good $I_C$ regions. For $(\delta_{dis},L_{dis})=(3.0\text{ meV},70\text{ nm})$ and 5000 steps of $\tilde{C}_C$, improved $C_C$ from $0.00 \rightarrow -0.053$. 
Fig. S3 shows the effect decreasing $L_{dis}$ has on our method. At 2.5 meV successful disorder mitigation of  $L_{dis}=50$ nm is achieved, though with rapid decay in functionality for smaller $L_{dis}$. $\delta_{dis}$ determines how small $L_{dis}$ can go while the mitigation method remains effective. For $\delta_{dis} = 3.0\text{ meV}$, even $L_{dis} = 60\text{ nm}$ is sufficient for $\tilde{C}_C$ mitigation to fail. While at  $L_{dis} = 30\text{ nm}$ for $\delta_{dis} = 1.5\text{ meV}$, $C_C$ goes from $0.0 \rightarrow -0.025$ using only $T_V$ mitigation and improves to $C_C = -0.044$ with a two-step process. The method failing at smaller $L_{dis}$ is not surprising, since more gates are needed affect shorter-length patterns. A comprehensive assessment can be found in Table S1.

\textit{Conclusion.\textemdash } Our disorder mitigation method, relying solely on conductance measurements, remarkably mitigates the detrimental effects of disorder suppressing topology in a Majorana nanowire. It is capable of transforming wires with no topologically non-trivial regions (in $I_C$ and $T_V$) into ones with robust topological passing regions across all stringent indicators. This is achieved using an exceptionally small "lite" neural network, derived from the larger, more robust vision transformer model introduced in our prior work \cite{taylor2025vision}. {We note that since we solve the Majorana Hamiltonian including disorder exactly, there are no additional quantum fluctuations to worry about in the problem.} We believe that our scheme can be directly applied to experimental Majorana nanowires \cite{aghaee2023inas}. {Using our protocol to experimental measurements should be straightforward as long as enough conductance data is collected in the measurements for the purpose of optimization.}

Despite the remarkable ability of our process to mitigate disorder, the "lite" neural network remains a significant limitation on its potential effectiveness. However, the smaller network offers one major advantage, the ability to use the larger network for verification. Specifically, if a wire is optimized using the "lite" model, the final gate voltages should not bias the larger neural network it does not interaction with. The mitigation procedure could be performed with the "lite" model while using the full model to determine the final phase diagram. The ability of the mitigation scheme to significantly improve the robustness of topological phases, especially when combined with the full model, can greatly reduce the search space, increasing the likelihood of success in subsequent braiding experiments. Braiding is essential for topological computation and the true test of topology is demonstrating this ability. 
The method's success is non-trivial, as the neural network is trained through supervised learning on a dataset that does not include deep steps of varying gate voltages required for optimization. It was possible the optimization could of failed due to the near-adversarial nature of applying an optimizer to a neural network of this type with potential for the network to hallucinate outside its domain of experience but it does not! We note, our method mitigates disorder rather than canceling it, reducing its impact on the wire's topology and thereby its usability for computation. One could explore modifying the neural network by training it to predict the effective magnitude of the disorder, utilizing it as a cost function to remove disorder directly. If the network accurately predicts total disorder, it may more aggressively reduce or even eliminate disorder in the system. Given that unintentional and unknown disorder is the key hindrance in achieving topological MZMs, our work should be useful in guiding future research on Majorana nanowires.

\textit{Acknowledgment.\textemdash } This work is supported by the Laboratory for Physical Sciences.  We thank Jay Deep Sau for many helpful discussions. We also thank UMD HPC Zaratan for computational resources provided. 
\putbib
\end{bibunit}
\begin{bibunit}
\clearpage
\pagebreak
\widetext
\begin{center}
\textbf{\large Supplemental Materials: Mitigating disorder and optimizing topological indicators with vision-transformer-based neural networks in Majorana nanowires}
\end{center}
\setcounter{equation}{0}
\setcounter{figure}{0}
\setcounter{table}{0}
\setcounter{page}{1}

\makeatletter
\renewcommand{\theequation}{S\arabic{equation}}
\renewcommand{\thefigure}{S\arabic{figure}}
\renewcommand{\thetable}{S\arabic{table}}
\renewcommand{\bibnumfmt}[1]{[S#1]}
\renewcommand{\citenumfont}[1]{S#1}
\setcounter{section}{0}
\renewcommand{\thesection}{S-\Roman{section}}
\makeatother
\section{Additional Results}

In this section we provide additional results showing the effectiveness of the disorder mitigation scheme for a variety of different correlation lengths and disorder magnitudes.
\begin{table}[h]
    \centering
    \begin{tabular}{|c|c|c|c|c|c|}
    \hline
        $\delta_{dis}$(meV)&$L_{dis}$(nm) &$C^i_{T_V}$ & $C^f_{T_V}$  & $C^i_{C}$ & $C^f_{C}$  \\
        \hline
        \multicolumn{6}{|c|}{Just $T_V$ Optimization}\\
        \hline
            0.000 & - & 0.155 & - & -0.287 & -\\
            1.500 & 70 & 0.292 & -0.127 & -0.006 & -0.107\\
            1.500 & 60 & 0.175 & -0.056 & -0.025 & -0.103\\ 
            1.500 & 50 & 0.714 & 0.020 & -0.025 & -0.083\\
            1.500 & 40 & 0.748 & 0.499 & 0.000 & -0.050\\
            1.500 & 30 & 0.660 & 0.294 & -0.009 & -0.025\\
            2.000 & 70 & 0.397 & -0.004 & -0.057 & -0.141\\
            2.000 & 60 & 0.831 & 0.237 & 0.000 & -0.057\\
            2.000 & 50 & 1.000 & 0.457 & 0.000 & -0.060\\
            2.500 & 70 & 0.961 & 0.388 & 0.000 & -0.077\\
            2.500 & 60 & 0.956 & 0.428 & 0.000 & -0.030\\
            2.500 & 50 & 1.000 & 0.717 & 0.000 & -0.015\\
            3.000 & 70 & 1.000 & 0.499 & 0.000 & 0.000\\ 
            3.000 & 60 & 1.000 & 0.912 & 0.000 & 0.000\\
            \hline
            \multicolumn{6}{|c|}{$I_c$ Mitigation}\\
            \hline
            1.500 & 30 & 0.660 & 0.184 & -0.009 & -0.044\\
            3.000 & 70 & 1.000 & 0.608 & 0.000 & -0.053\\
            3.000 & 60 & 1.000 & 0.881 & 0.000 & 0.000\\
            \hline
            
    \end{tabular}
    \caption{Summarization of results of the disorder mitigation algorithm for different disorder magnitudes $\delta_{dis}$ (meV) and disorder correlation lengths $L_{dis}$ (nm).  The columns $C^i_{T_V}$ and $C^f_{T_V}$ refer to the $T_V$ cost function before and after the procedure respectively, and similarly for $C^i_{C}$ and $C^f_{C}$ with the combined parameter $I_C$. The first block refers to the procedure performed for 2000 iterations using $C_{T_V}$ as the cost, while the second block refers to results making use of $\tilde{C}_C$ to optimize.}
    \label{tab:fullres}
\end{table} 
\subsection{Mitigation }
\begin{figure}[H]
     \centering
     \begin{subfigure}[b]{0.45\columnwidth}
         \centering
         \includegraphics[width=\textwidth]{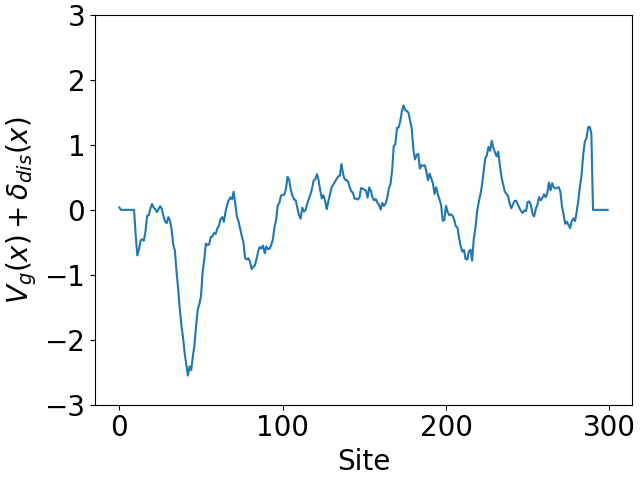}
         \caption{}
     \end{subfigure}
          \begin{subfigure}[b]{0.45\columnwidth}
         \centering
         \includegraphics[width=\textwidth]{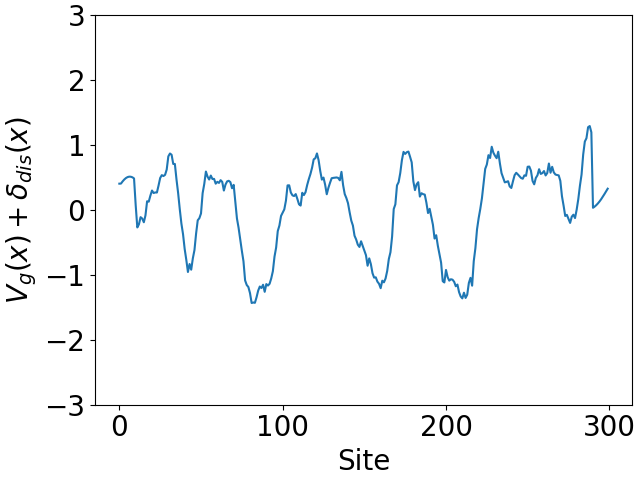}
         \caption{}
     \end{subfigure}
     \caption{$\delta_{dis}(x)+V_g(x)$ in $meV$ before (a) and after (b) mitigation for a wire with $\delta_{dis}=2.5$ meV and $L_{dis}=70$ nm. This is the same wire shown within the main text.}
     \label{fig:Cond}
\end{figure}

\begin{figure}[H]
     \centering
     \begin{subfigure}[b]{0.7\columnwidth}
         \centering
         \includegraphics[width=\textwidth]{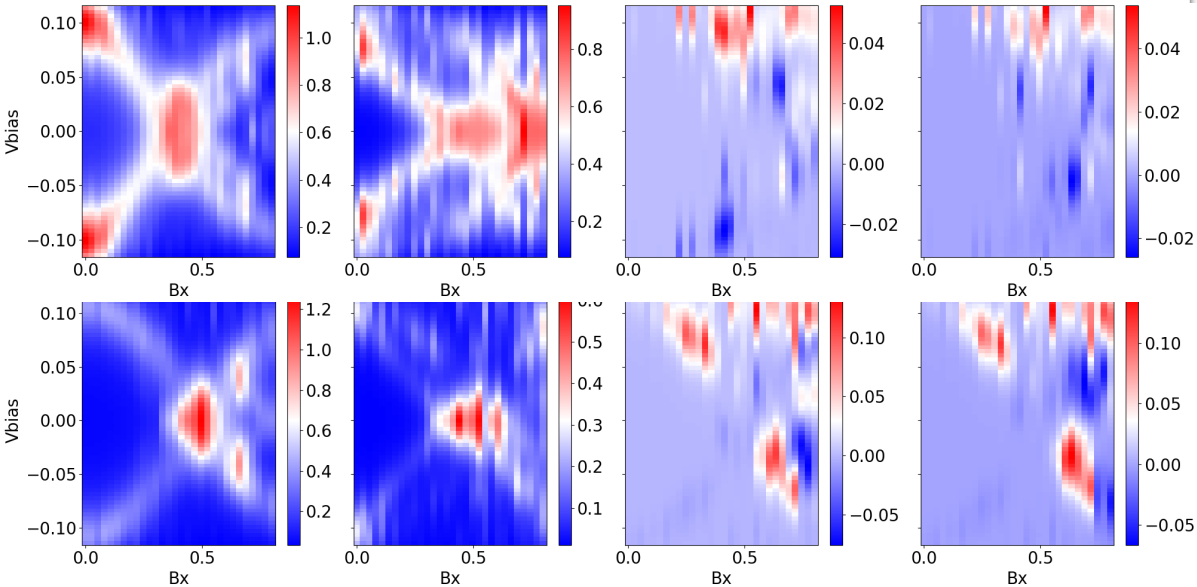}
         \caption{}
     \end{subfigure}
          \begin{subfigure}[b]{0.7\columnwidth}
         \centering
         \includegraphics[width=\textwidth]{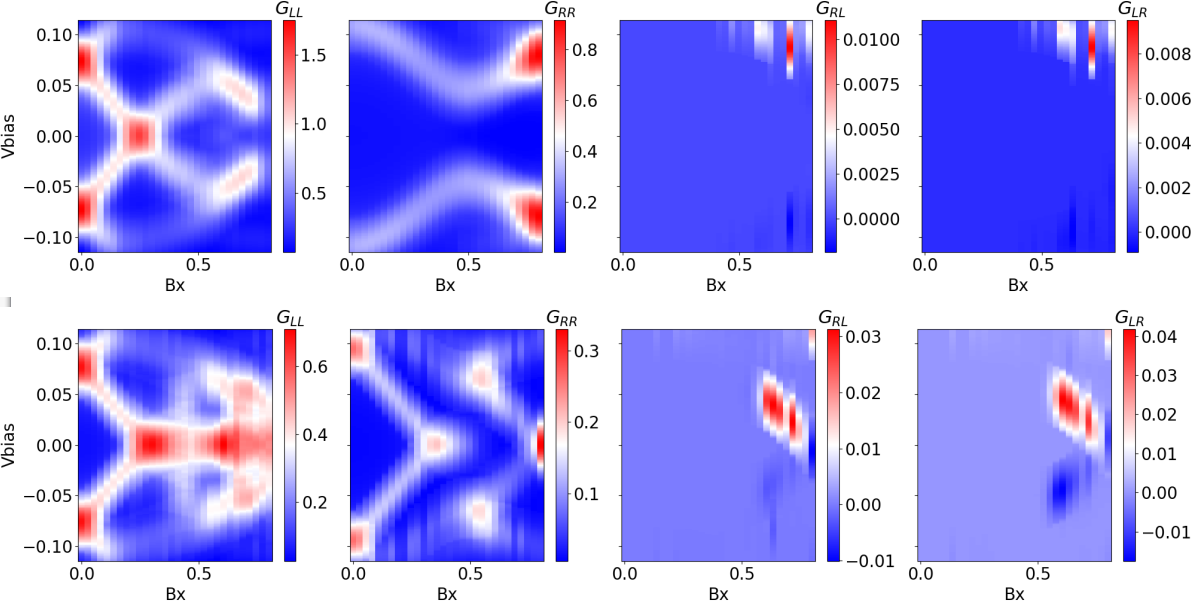}
         \caption{}
     \end{subfigure}
     \caption{Conductance results plotted against $V_{Bias}$ and B for a wire realization at $\mu,\delta_{dis},L_{dis}$ = (a) (0.4meV, 2.5meV, 70nm) and (b) (0.3meV, 3.0meV, 60nm). Results are shown for a successful and unsuccessful cases of mitigation in (a) and (b) respectively. In (a-b) The first and second rows consists of conductances before and after the mitigation scheme respectively.}
     \label{fig:discompare}
\end{figure}

Within this section, we provide additional results to illustrate that the mitigation scheme does not cancel out disorder but mitigates its effect on the topology of a wire. In Fig. \ref{fig:discompare}, we present the effective disorder potential ($\delta_{dis}(x)+V_g(x)$) for the 2.5 meV wire shown in Fig. \ref{fig:2,5BAIndicators} and \ref{fig:timeresults} in the main text. Despite significant improvements in topology, there is no clear evidence of a substantial reduction in disorder. In Fig. \ref{fig:Cond}, we provide examples of the differential conductance for cases of both successful and unsuccessful mitigation. The conductance, which appears to exhibit seemingly high disorder in both cases, serves to emphasize the unique approach the algorithm takes in mitigating disorder. In particular, the conductance data for the 2.5 meV wire in Fig. \ref{fig:Cond} show that post-mitigation conductance does not strongly resemble that of a low-disorder wire, despite $T_V$ and $I_C$ indicating that the wire passes topological tests. This highlights that the mitigation method does not remove disorder but instead modifies it to reduce its impact on $T_V$ in a way that can not be predicted before utilization.

\newpage
\subsection{$T_V$ and $I_C$ Results}
 Below, we provide additional plots showing the improvement of the scattering invariant $T_V$ along with the combined indicator $I_C$. We provide the plots for results referenced in Table \ref{tab:fullres}. Results are presented for a wide variety of different $\delta_{dis}$ and $L_{dis}$ along with some for combinations of different iteration types, mitigating either for $C_{T_V}$ or $C_C$. It should be noted that $C_C$ and $C_{T_V}$ are computed at the resolution of the neural network (though not using the neural network), whereas these plots have a much higher resolution.

\begin{figure}[h]
     \centering
     \begin{subfigure}[b]{0.9\columnwidth}
         \centering
         \includegraphics[width=\textwidth]{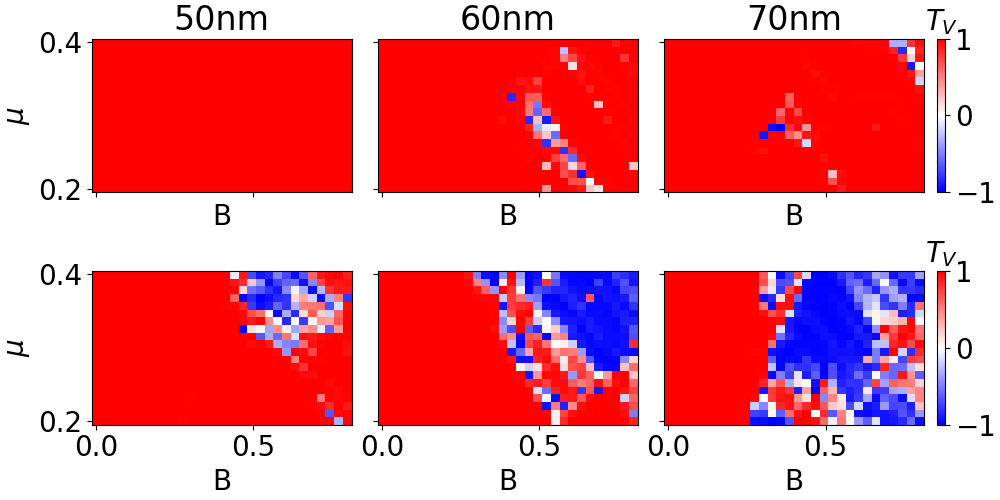}
         \caption{}
     \end{subfigure}
     \caption{$T_V$ plotted against $\mu$ and B for representative wire realizations of different $L_{dis}$ and fixed $\delta_{dis}=2.50$meV. The first and second rows consists of $T_V$ before and after the mitigation scheme respectively.}
     \label{fig:LenComp}
\end{figure}
\begin{figure}[H]
     \centering
         \begin{subfigure}[b]{0.7\columnwidth}
         \centering
         \includegraphics[width=\textwidth]{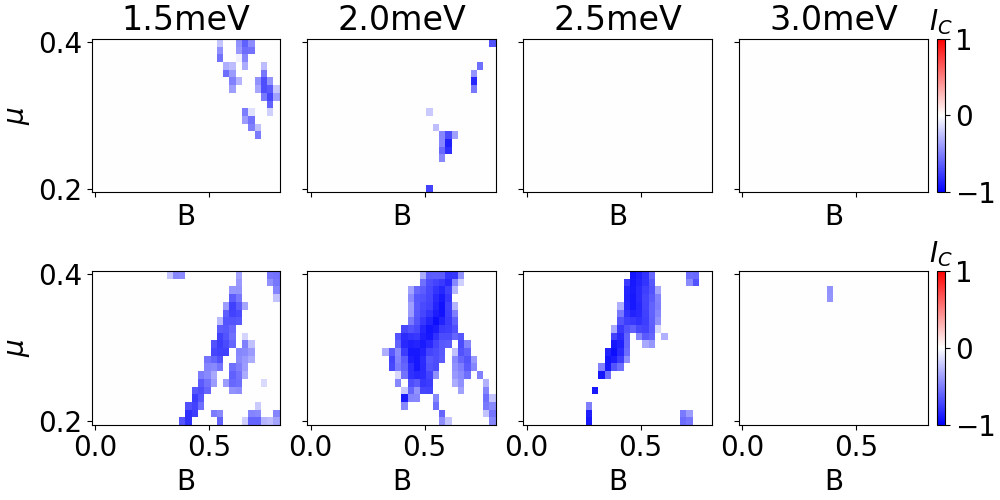}
         \caption{}
     \end{subfigure}
     \caption{$I_C$ plotted against $\mu$ and B for representative wire realizations of different disorder magnitudes and fixed $L_{dis}=70$nm. The first and second rows consists of $I_C$ before and after the mitigation scheme respectively. These plots are for the mitigation scheme optimizing only $C_{T_V}$ with 2000 evaluations.}
     \label{fig:S_60nm}
\end{figure}

\begin{figure}[H]
     \centering
     \begin{subfigure}[b]{0.7\columnwidth}
         \centering
         \includegraphics[width=\textwidth]{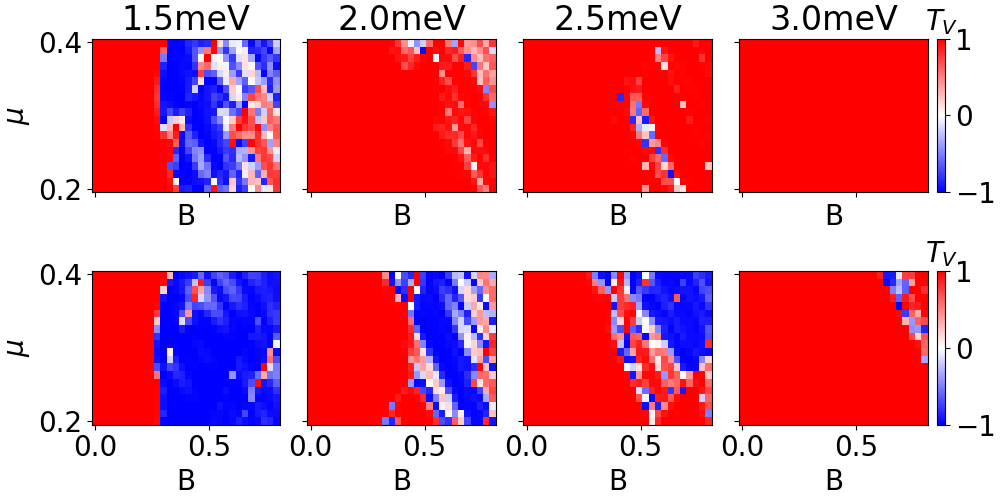}
         \caption{}
     \end{subfigure}
         \begin{subfigure}[b]{0.7\columnwidth}
         \centering
         \includegraphics[width=\textwidth]{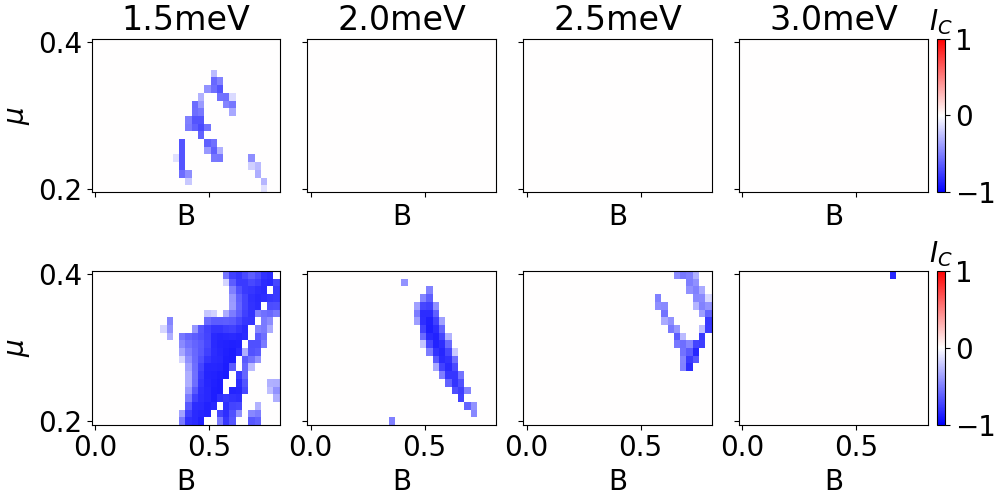}
         \caption{}
     \end{subfigure}
     \caption{(a) $T_V$ and (b) $I_C$ plotted against $\mu$ and B for representative wire realizations of different disorder magnitudes and fixed $L_{dis}=60$nm. The first and second rows consists of $T_V$ before and after the mitigation scheme respectively. These plots are for the mitigation scheme optimizing only $C_{T_V}$ with 2000 evaluations.}
     \label{fig:S_60nm}
\end{figure}

\begin{figure}[H]
     \centering
     \begin{subfigure}[b]{0.7\columnwidth}
         \centering
         \includegraphics[width=\textwidth]{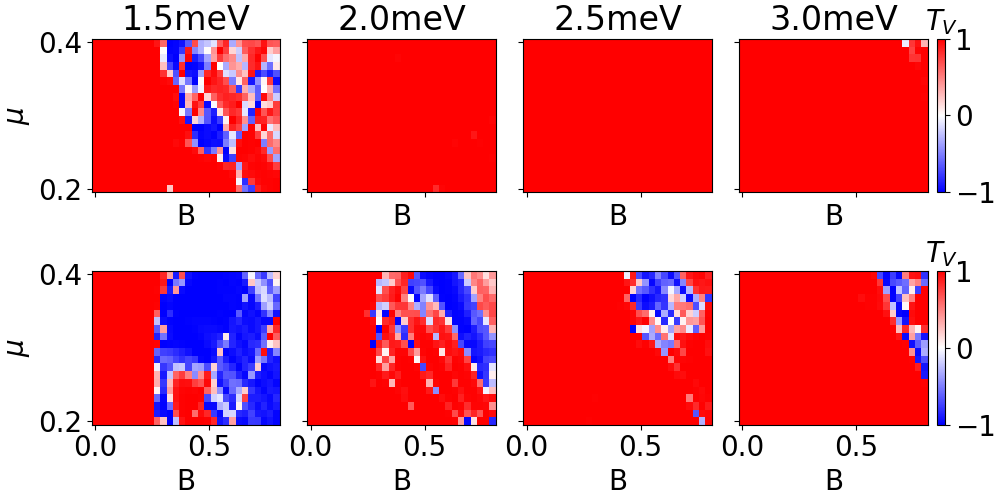}
         \caption{}
     \end{subfigure}
          \begin{subfigure}[b]{0.7\columnwidth}
         \centering
         \includegraphics[width=\textwidth]{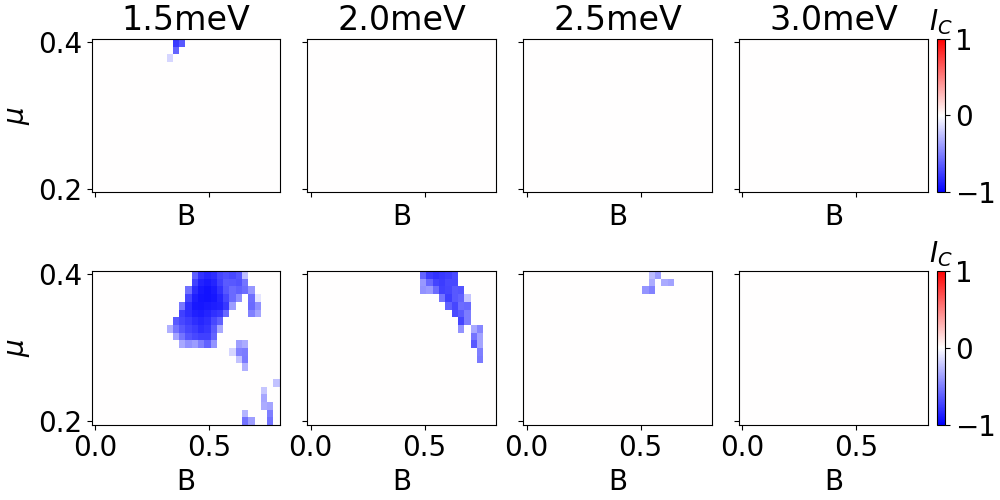}
         \caption{}
     \end{subfigure}
     \caption{(a) $T_V$ and (b) $I_C$ plotted against $\mu$ and B for representative wire realizations of different disorder magnitudes and fixed $L_{dis}=50$nm. The first and second rows consists of (a) $T_V$ and (b) $I_C$ before and after the mitigation scheme respectively.These plots are for the mitigation scheme optimizing only $C_{T_V}$ with 2000 evaluations.}
     \label{fig:S_50nm}
\end{figure}

\begin{figure}[H]
     \centering
     \begin{subfigure}[b]{0.7\columnwidth}
         \centering
         \includegraphics[width=\textwidth]{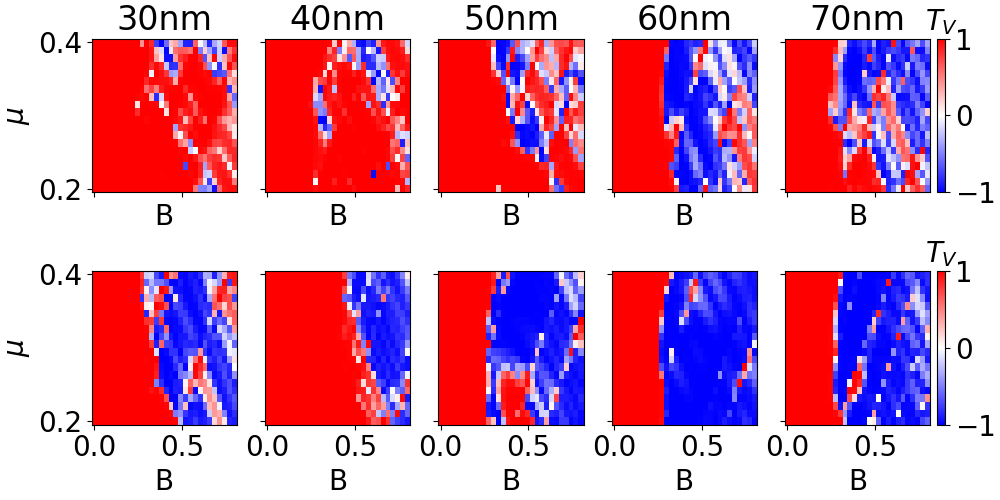}
         \caption{}
     \end{subfigure}
          \begin{subfigure}[b]{0.7\columnwidth}
         \centering
         \includegraphics[width=\textwidth]{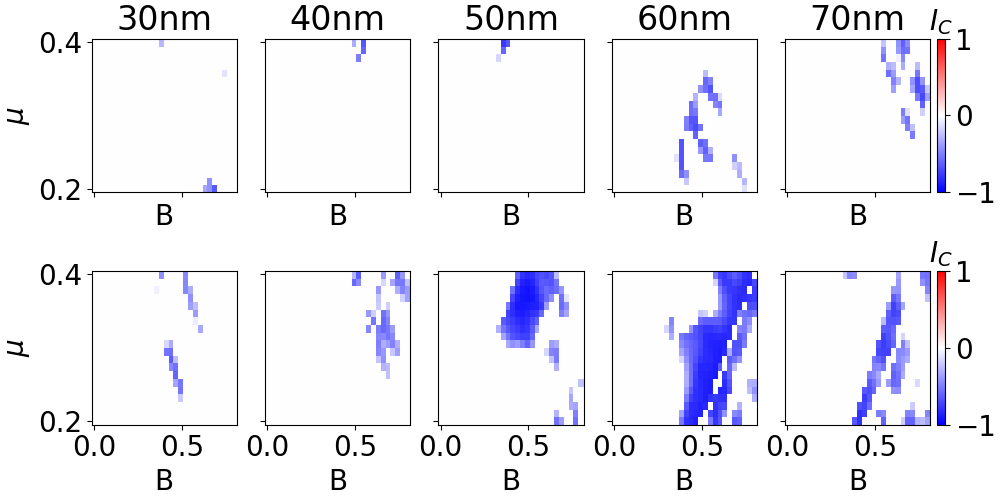}
         \caption{}
     \end{subfigure}
     \caption{(a) $T_V$ and (b) $I_C$ plotted against $\mu$ and B for representative wire realizations of different $L_{dis}$ and fixed $\delta_{dis}=1.5$meV. The first and second rows consists of (a) $T_V$ and (b) $I_C$ before and after the mitigation scheme respectively. The mitigation in this plot was ran only for $C_{T_V}$ without additional $I_C$ mitigation.}
     \label{fig:S_1.5meV}
\end{figure}

\begin{figure}[H]
     \centering
     \begin{subfigure}[b]{0.45\columnwidth}
         \centering
         \includegraphics[width=\textwidth]{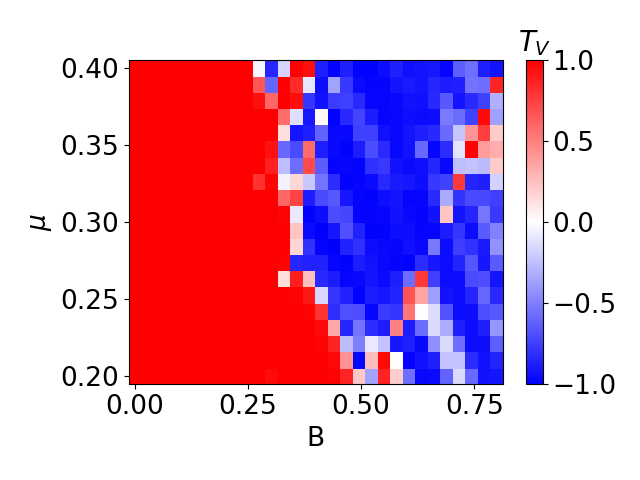}
         \caption{}
     \end{subfigure}
          \begin{subfigure}[b]{0.45\columnwidth}
         \centering
         \includegraphics[width=\textwidth]{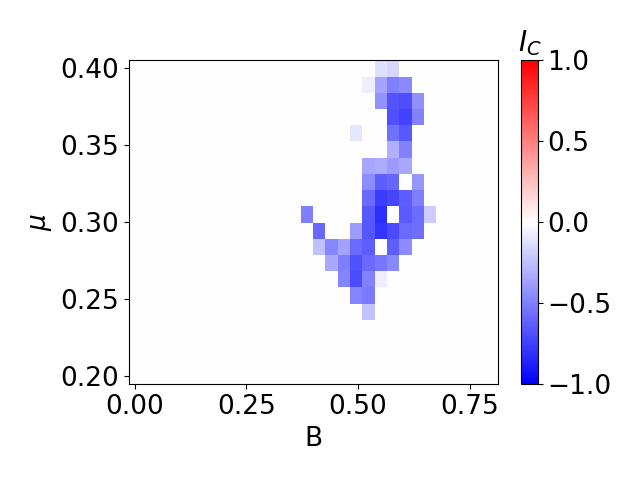}
         \caption{}
     \end{subfigure}
     \caption{Post follow-up $\tilde{C}_C$ mitigation scheme (a) $T_V$ and (b) $I_C$ plotted against $\mu$ and B for representative wire realization of $L_{dis}=30nm$ and $\delta_{dis}=1.5$meV. The follow-up mitigation optimizing for $\tilde{C}_{C}$ was performed for 1000 iterations continuing from the result shown in Fig. \ref{fig:S_1.5meV}.}
     \label{fig:S_1.5meVFollowup}
\end{figure}

\begin{figure}[H]
     \centering
     \begin{subfigure}[b]{0.45\columnwidth}
         \centering
         \includegraphics[width=\textwidth]{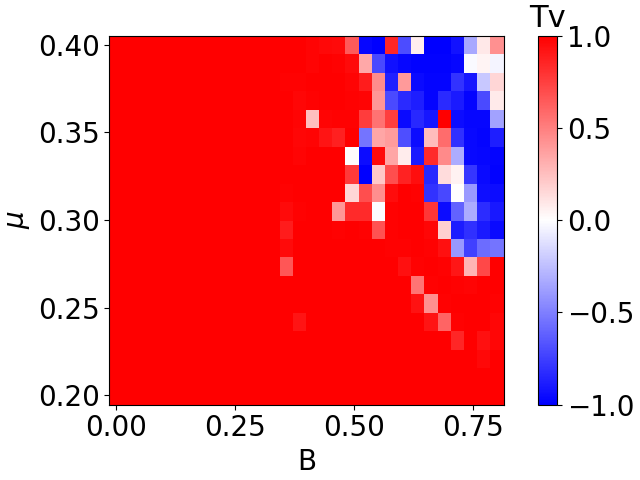}
         \caption{}
     \end{subfigure}
          \begin{subfigure}[b]{0.45\columnwidth}
         \centering
         \includegraphics[width=\textwidth]{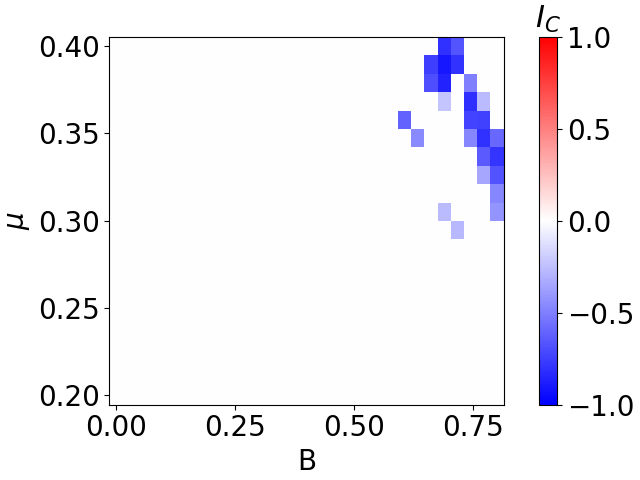}
         \caption{}
     \end{subfigure}
     \caption{$\tilde{C}_C$ mitigation scheme (a) $T_V$ and (b) $I_C$ plotted against $\mu$ and B for representative wire realization of $L_{dis}=70nm$ and $\delta_{dis}=3.0$meV. The $\tilde{C}_C$ mitigation was ran for 5000 iterations with random initialization.}
     \label{fig:S_1.5meVFollowup}
\end{figure}

\section{Additional Machine Learning details}
\label{sec:addml}

\subsection{Neural Network Details}
\label{sec:neuralnetworkdetails}
The neural network is based on the network introduced within \cite{taylor2025vision} employing a vision transformer-based architecture \cite{dosovitskiy2020image} with a 3D convolutional layer for initial patching. As a result, the input is a three-dimensional matrix corresponding to the three varied experimental parameters, with four "color" channels representing the four differential conductances. The input consists of conductance measurements for 3 values of $\mu$, 10 values of $B$, and 31 values of $V_{Bias}$, with $\mu \in [0.2\text{meV}, 0.4\text{meV}]$, $B \in [0 T, 0.8T]$, and $V_{Bias} \in [-0.05\text{meV}, +0.05\text{meV}]$. This reduced number of measurements results in a "lite" version of the full topological prediction network \cite{taylor2025vision}, although significant improvements are likely if greater computational resources allow the use of the full version. 

\begin{figure}[H]
     \centering
     \begin{subfigure}[b]{0.30\columnwidth}
         \centering
         \includegraphics[width=\textwidth]{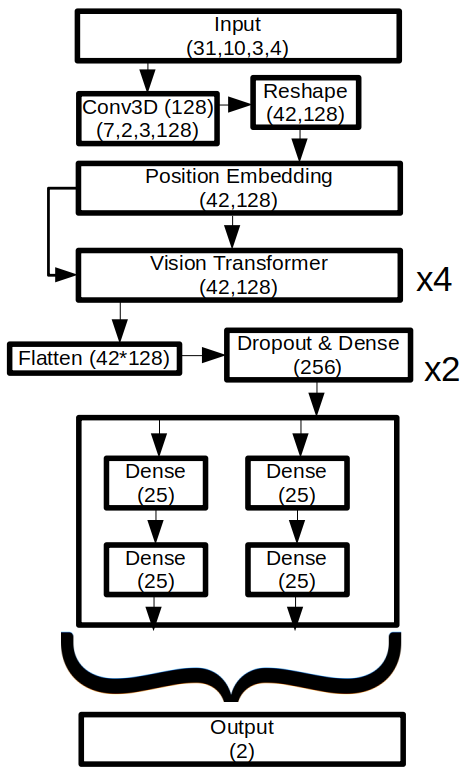}
         \caption{}
     \end{subfigure}
     \begin{subfigure}[b]{0.3\columnwidth}
         \centering
         \includegraphics[width=\textwidth]{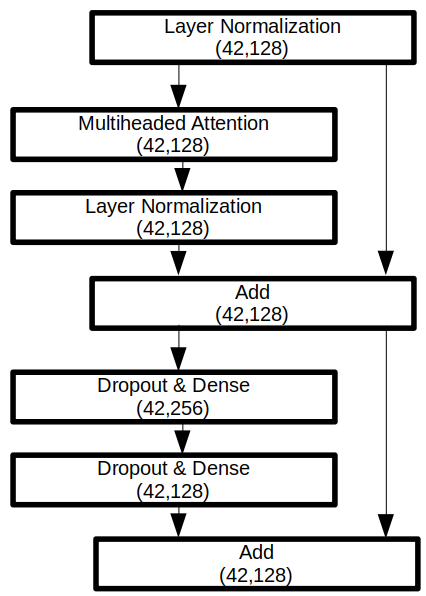}
         \caption{}
     \end{subfigure}
     \caption{(a) Diagram of the neural network used to process conductance measurements into predictions of topological/Majorana indicators. The neural network used is similar to standard vision transformer-based models, except with a many-path tree ending, allowing $C_{T_V}$, $C_C$  to undergo some minor independent processing. Given the small number of outputs within this neural network, these independent branches are likely not necessary.  Further details about the neural network's architecture can be found in the main text. (b) Diagram of the vision transformer used within (a). The vision transformer works by first performing layer normalization on the data input. It then has two additional paths: a skip path, which is used to prevent the vanishing gradient problem, and the multi-head attention path (followed by layer normalization). These two paths are combined additively. After this, another two paths form: the first where the MLP is used with dropout, consisting of two dense layers with 256 and 128 nodes, and the other another skip layer. These two paths are then also combined additively. }
     \label{fig:NNDiagram}
\end{figure}

The output differs from previous work; instead of predicting the entire phase diagram, which is challenging given the low-resolution input in the measurement matrix, the two cost functions, $C_C(\gamma=0)$ and $C_{T_V}$, are directly produced and can be predicted with high accuracy.  This is possible because the complex patterns in the disordered phase diagram, which make direct phase prediction difficult, are lost when averaging in the cost function. This is implemented by making use of only 2 branches (instead of the previous 100 where each branch was for a specific phase point), namely one for $C_{T_V}$ and one for $C_{C}$. When combined with the decreased number of measurements this change significantly reduces the size of the neural network. A diagram of the neural network can be seen in Fig. \ref{fig:NNDiagram}.

The training data consisted of a large set of device realizations with varying gate voltage configurations. Specifically, the dataset included cases with all gates set to zero, cases with random gate voltages, and cases where a few steps of gradient descent were applied to set $V_g$ such that $\delta_{dis}(x) + V_g(x) = 0$, simulating partial optimization of a larger disorder. This approach helped the neural network more accurately predict the indicators in the presence of non-zero gate voltages. The neural network is trained with a fixed number of gates, $N_{g} = 20$, where $\beta_j$ magnitudes are set between [-1.0, 1.0], with a bias toward [-0.5, 0.5]. A total of 339,999 devices were simulated, with low computational cost due to the small model. In testing, the network accurately predicted $C_{T_V}$ and $C_C$, with errors of $\Delta C_{T_V}=0.0374$ and $\Delta C_{C} = 0.0087$. The parameters $\delta_{dis}$, $L_{dis}$, and $\alpha$ were randomly sampled from the full distribution in \cite{taylor2025vision}, using a single neural network for the entire range.

\subsection{Training Data Generation}
This section provides a more in-depth explanation of the training data generation process. The process is based on the archetype described in \cite{taylor2024machine,taylor2024neural,taylor2025vision}. The training data is generated by simulating various 1D nanowire devices, each with randomly assigned disorder properties. Each device is characterized by a "disorder vector" $\vec{D}$, which includes disorder potentials, spin-orbit coupling strength, correlation length, and disorder magnitude. Experimental parameters such as chemical potential $\mu$ and magnetic field $B$ are recorded in a matrix $\textbf{K}$. $\textbf{K}$ is constant thus the same measurements are performed for every device and voltage characterization $\vec{\beta}$. For each measurement configuration (a row of $ \textbf{K} $), conductance values are collected into an input matrix $\textbf{X}$ which when fed into the neural network is reshaped such that its axis correspond to conductances for different values of $V_{Bias}, \mu, B$:
$$
X^j = [G_{LL}^j, G_{LR}^j, G_{RL}^j, G_{RR}^j]
$$
The differential conductances are defined as $G_{\alpha \beta}= dI_\alpha/dV_\beta$ for $\alpha =\beta$ and $G_{\alpha \beta}= -dI_\alpha/dV_\beta$ for $\alpha \neq \beta$. The neural network is trained to predict an output vector $\textbf{Y}$, which contains $C_{T_V}$ and $C_{C}$ for a particular device realization and gate voltage characterization $\vec{\beta}$:
$$
Y = \begin{bmatrix} C_{T_V} \\ C_{C} \end{bmatrix}
$$
The training data generator function, which is calculated through KWANT \cite{groth2014kwant}, for device $i$ is given by:
$$
f_\text{Gen} (\vec{D}_i, \vec{\beta},\textbf{K}) = [\textbf{X}_i, \textbf{Y}_i]
$$
The neural network then learns to predict $\vec{Y}_i$ from $\textbf{X}_i$ and $\textbf{K}$:
$$
f_\text{NN}(\textbf{X}_i, \textbf{K}) = \vec{Y}_i
$$
It should be noted that the neural network is not given access to the values of $\vec{\beta}$, this was done with the intention of forcing it to actually learn the features of the conductance data instead of relying $\vec{\beta}$ which makes it more robust. This also handles a problem with the training data where large random gate potentials almost always just harm topology, so its likely the neural network which could lead to undesirable hallucinations. The approach outlined here allows the model to determine our indicators using only conductance measurements, without requiring knowledge of the underlying disorder parameters, making it directly applicable to experiments.

\subsection{Algorithm}
\label{sec:cmaesdetails}
The mitigation scheme made use of CMA-ES \cite{hansen2016cma} to perform the gate optimization steps. CMA-ES was chosen due to it both being used within prior work and for its ability to handle complex non-differentiable optimization landscapes. There are many alternative optimization algorithms that may be more effective but we did not assess them. We made use of the fcmaes package \cite{fcmaes2021} to allow for parallel evaluation of the entire population at once while setting our population size $n_{pop}=15$ which is quite limited due to the core limitations. The population sigma was initially set to 0.1 and automatically adjusts as more iterations are performed. This was true both in initial $C_{T_V}$ optimizations and follow-up $C_C$. During the optimization the bottleneck was entirely within the KWANT data generation, a bottleneck that would not exist if our method was applied to experiment. In particular our optimization procedure made use of a queuing system where a single instance of the neural network was loaded at a time doing all $C^{pred}_{C}$ and $C^{pred}_{T_V}$ calculations and this did not limit throughput at all. We placed bounds on $\vec{\beta}$ at $[-0.5,0.5]$ in most cases, however in cases requiring further optimization we switched to $[-1,1]$. We did not notice any significant difference in results by changing these bounds.

\putbib
\end{bibunit}


\begin{thebibliography}{41}%
\makeatletter
\providecommand \@ifxundefined [1]{%
 \@ifx{#1\undefined}
}%
\providecommand \@ifnum [1]{%
 \ifnum #1\expandafter \@firstoftwo
 \else \expandafter \@secondoftwo
 \fi
}%
\providecommand \@ifx [1]{%
 \ifx #1\expandafter \@firstoftwo
 \else \expandafter \@secondoftwo
 \fi
}%
\providecommand \natexlab [1]{#1}%
\providecommand \enquote  [1]{``#1''}%
\providecommand \bibnamefont  [1]{#1}%
\providecommand \bibfnamefont [1]{#1}%
\providecommand \citenamefont [1]{#1}%
\providecommand \href@noop [0]{\@secondoftwo}%
\providecommand \href [0]{\begingroup \@sanitize@url \@href}%
\providecommand \@href[1]{\@@startlink{#1}\@@href}%
\providecommand \@@href[1]{\endgroup#1\@@endlink}%
\providecommand \@sanitize@url [0]{\catcode `\\12\catcode `\$12\catcode `\&12\catcode `\#12\catcode `\^12\catcode `\_12\catcode `\%12\relax}%
\providecommand \@@startlink[1]{}%
\providecommand \@@endlink[0]{}%
\providecommand \url  [0]{\begingroup\@sanitize@url \@url }%
\providecommand \@url [1]{\endgroup\@href {#1}{\urlprefix }}%
\providecommand \urlprefix  [0]{URL }%
\providecommand \Eprint [0]{\href }%
\providecommand \doibase [0]{https://doi.org/}%
\providecommand \selectlanguage [0]{\@gobble}%
\providecommand \bibinfo  [0]{\@secondoftwo}%
\providecommand \bibfield  [0]{\@secondoftwo}%
\providecommand \translation [1]{[#1]}%
\providecommand \BibitemOpen [0]{}%
\providecommand \bibitemStop [0]{}%
\providecommand \bibitemNoStop [0]{.\EOS\space}%
\providecommand \EOS [0]{\spacefactor3000\relax}%
\providecommand \BibitemShut  [1]{\csname bibitem#1\endcsname}%
\let\auto@bib@innerbib\@empty
\bibitem [{\citenamefont {Lutchyn}\ \emph {et~al.}(2010)\citenamefont {Lutchyn}, \citenamefont {Sau},\ and\ \citenamefont {Das~Sarma}}]{lutchyn2010majorana}%
  \BibitemOpen
  \bibfield  {author} {\bibinfo {author} {\bibfnamefont {R.~M.}\ \bibnamefont {Lutchyn}}, \bibinfo {author} {\bibfnamefont {J.~D.}\ \bibnamefont {Sau}},\ and\ \bibinfo {author} {\bibfnamefont {S.}~\bibnamefont {Das~Sarma}},\ }\href@noop {} {\bibfield  {journal} {\bibinfo  {journal} {Physical review letters}\ }\textbf {\bibinfo {volume} {105}},\ \bibinfo {pages} {077001} (\bibinfo {year} {2010})}\BibitemShut {NoStop}%
\bibitem [{\citenamefont {Sau}\ \emph {et~al.}(2010{\natexlab{a}})\citenamefont {Sau}, \citenamefont {Lutchyn}, \citenamefont {Tewari},\ and\ \citenamefont {Das~Sarma}}]{sau2010robustness}%
  \BibitemOpen
  \bibfield  {author} {\bibinfo {author} {\bibfnamefont {J.~D.}\ \bibnamefont {Sau}}, \bibinfo {author} {\bibfnamefont {R.~M.}\ \bibnamefont {Lutchyn}}, \bibinfo {author} {\bibfnamefont {S.}~\bibnamefont {Tewari}},\ and\ \bibinfo {author} {\bibfnamefont {S.}~\bibnamefont {Das~Sarma}},\ }\href@noop {} {\bibfield  {journal} {\bibinfo  {journal} {Physical Review B—Condensed Matter and Materials Physics}\ }\textbf {\bibinfo {volume} {82}},\ \bibinfo {pages} {094522} (\bibinfo {year} {2010}{\natexlab{a}})}\BibitemShut {NoStop}%
\bibitem [{\citenamefont {Oreg}\ \emph {et~al.}(2010)\citenamefont {Oreg}, \citenamefont {Refael},\ and\ \citenamefont {Von~Oppen}}]{oreg2010helical}%
  \BibitemOpen
  \bibfield  {author} {\bibinfo {author} {\bibfnamefont {Y.}~\bibnamefont {Oreg}}, \bibinfo {author} {\bibfnamefont {G.}~\bibnamefont {Refael}},\ and\ \bibinfo {author} {\bibfnamefont {F.}~\bibnamefont {Von~Oppen}},\ }\href@noop {} {\bibfield  {journal} {\bibinfo  {journal} {Physical review letters}\ }\textbf {\bibinfo {volume} {105}},\ \bibinfo {pages} {177002} (\bibinfo {year} {2010})}\BibitemShut {NoStop}%
\bibitem [{\citenamefont {Sau}\ \emph {et~al.}(2010{\natexlab{b}})\citenamefont {Sau}, \citenamefont {Lutchyn}, \citenamefont {Tewari},\ and\ \citenamefont {Das~Sarma}}]{sau2010generic}%
  \BibitemOpen
  \bibfield  {author} {\bibinfo {author} {\bibfnamefont {J.~D.}\ \bibnamefont {Sau}}, \bibinfo {author} {\bibfnamefont {R.~M.}\ \bibnamefont {Lutchyn}}, \bibinfo {author} {\bibfnamefont {S.}~\bibnamefont {Tewari}},\ and\ \bibinfo {author} {\bibfnamefont {S.}~\bibnamefont {Das~Sarma}},\ }\href@noop {} {\bibfield  {journal} {\bibinfo  {journal} {Physical review letters}\ }\textbf {\bibinfo {volume} {104}},\ \bibinfo {pages} {040502} (\bibinfo {year} {2010}{\natexlab{b}})}\BibitemShut {NoStop}%
\bibitem [{\citenamefont {Das~Sarma}(2023)}]{das2023search}%
  \BibitemOpen
  \bibfield  {author} {\bibinfo {author} {\bibfnamefont {S.}~\bibnamefont {Das~Sarma}},\ }\href@noop {} {\bibfield  {journal} {\bibinfo  {journal} {Nature Physics}\ }\textbf {\bibinfo {volume} {19}},\ \bibinfo {pages} {165} (\bibinfo {year} {2023})}\BibitemShut {NoStop}%
\bibitem [{\citenamefont {Sarma}\ \emph {et~al.}(2015)\citenamefont {Sarma}, \citenamefont {Freedman},\ and\ \citenamefont {Nayak}}]{sarma2015majorana}%
  \BibitemOpen
  \bibfield  {author} {\bibinfo {author} {\bibfnamefont {S.~D.}\ \bibnamefont {Sarma}}, \bibinfo {author} {\bibfnamefont {M.}~\bibnamefont {Freedman}},\ and\ \bibinfo {author} {\bibfnamefont {C.}~\bibnamefont {Nayak}},\ }\href@noop {} {\bibfield  {journal} {\bibinfo  {journal} {npj Quantum Information}\ }\textbf {\bibinfo {volume} {1}},\ \bibinfo {pages} {1} (\bibinfo {year} {2015})}\BibitemShut {NoStop}%
\bibitem [{\citenamefont {Lutchyn}\ \emph {et~al.}(2018)\citenamefont {Lutchyn}, \citenamefont {Bakkers}, \citenamefont {Kouwenhoven}, \citenamefont {Krogstrup}, \citenamefont {Marcus},\ and\ \citenamefont {Oreg}}]{lutchyn2018majorana}%
  \BibitemOpen
  \bibfield  {author} {\bibinfo {author} {\bibfnamefont {R.~M.}\ \bibnamefont {Lutchyn}}, \bibinfo {author} {\bibfnamefont {E.~P.}\ \bibnamefont {Bakkers}}, \bibinfo {author} {\bibfnamefont {L.~P.}\ \bibnamefont {Kouwenhoven}}, \bibinfo {author} {\bibfnamefont {P.}~\bibnamefont {Krogstrup}}, \bibinfo {author} {\bibfnamefont {C.~M.}\ \bibnamefont {Marcus}},\ and\ \bibinfo {author} {\bibfnamefont {Y.}~\bibnamefont {Oreg}},\ }\href@noop {} {\bibfield  {journal} {\bibinfo  {journal} {Nature Reviews Materials}\ }\textbf {\bibinfo {volume} {3}},\ \bibinfo {pages} {52} (\bibinfo {year} {2018})}\BibitemShut {NoStop}%
\bibitem [{\citenamefont {Kitaev}(2001)}]{kitaev2001unpaired}%
  \BibitemOpen
  \bibfield  {author} {\bibinfo {author} {\bibfnamefont {A.~Y.}\ \bibnamefont {Kitaev}},\ }\href@noop {} {\bibfield  {journal} {\bibinfo  {journal} {Physics-uspekhi}\ }\textbf {\bibinfo {volume} {44}},\ \bibinfo {pages} {131} (\bibinfo {year} {2001})}\BibitemShut {NoStop}%
\bibitem [{\citenamefont {Ahn}\ \emph {et~al.}(2021)\citenamefont {Ahn}, \citenamefont {Pan}, \citenamefont {Woods}, \citenamefont {Stanescu},\ and\ \citenamefont {Sarma}}]{ahn2021estimating}%
  \BibitemOpen
  \bibfield  {author} {\bibinfo {author} {\bibfnamefont {S.}~\bibnamefont {Ahn}}, \bibinfo {author} {\bibfnamefont {H.}~\bibnamefont {Pan}}, \bibinfo {author} {\bibfnamefont {B.}~\bibnamefont {Woods}}, \bibinfo {author} {\bibfnamefont {T.~D.}\ \bibnamefont {Stanescu}},\ and\ \bibinfo {author} {\bibfnamefont {S.~D.}\ \bibnamefont {Sarma}},\ }\href@noop {} {\bibfield  {journal} {\bibinfo  {journal} {Physical Review Materials}\ }\textbf {\bibinfo {volume} {5}},\ \bibinfo {pages} {124602} (\bibinfo {year} {2021})}\BibitemShut {NoStop}%
\bibitem [{\citenamefont {Woods}\ \emph {et~al.}(2021)\citenamefont {Woods}, \citenamefont {Sarma},\ and\ \citenamefont {Stanescu}}]{woods2021charge}%
  \BibitemOpen
  \bibfield  {author} {\bibinfo {author} {\bibfnamefont {B.~D.}\ \bibnamefont {Woods}}, \bibinfo {author} {\bibfnamefont {S.~D.}\ \bibnamefont {Sarma}},\ and\ \bibinfo {author} {\bibfnamefont {T.~D.}\ \bibnamefont {Stanescu}},\ }\href@noop {} {\bibfield  {journal} {\bibinfo  {journal} {Physical Review Applied}\ }\textbf {\bibinfo {volume} {16}},\ \bibinfo {pages} {054053} (\bibinfo {year} {2021})}\BibitemShut {NoStop}%
\bibitem [{\citenamefont {Kitaev}(2003)}]{kitaev2003fault}%
  \BibitemOpen
  \bibfield  {author} {\bibinfo {author} {\bibfnamefont {A.~Y.}\ \bibnamefont {Kitaev}},\ }\href@noop {} {\bibfield  {journal} {\bibinfo  {journal} {Annals of Physics}\ }\textbf {\bibinfo {volume} {303}},\ \bibinfo {pages} {2} (\bibinfo {year} {2003})}\BibitemShut {NoStop}%
\bibitem [{\citenamefont {Sengupta}\ \emph {et~al.}(2001)\citenamefont {Sengupta}, \citenamefont {{\v{Z}}uti{\'c}}, \citenamefont {Kwon}, \citenamefont {Yakovenko},\ and\ \citenamefont {Sarma}}]{sengupta2001midgap}%
  \BibitemOpen
  \bibfield  {author} {\bibinfo {author} {\bibfnamefont {K.}~\bibnamefont {Sengupta}}, \bibinfo {author} {\bibfnamefont {I.}~\bibnamefont {{\v{Z}}uti{\'c}}}, \bibinfo {author} {\bibfnamefont {H.-J.}\ \bibnamefont {Kwon}}, \bibinfo {author} {\bibfnamefont {V.~M.}\ \bibnamefont {Yakovenko}},\ and\ \bibinfo {author} {\bibfnamefont {S.~D.}\ \bibnamefont {Sarma}},\ }\href@noop {} {\bibfield  {journal} {\bibinfo  {journal} {Physical Review B}\ }\textbf {\bibinfo {volume} {63}},\ \bibinfo {pages} {144531} (\bibinfo {year} {2001})}\BibitemShut {NoStop}%
\bibitem [{\citenamefont {Pan}\ and\ \citenamefont {Sarma}(2020)}]{pan2020physical}%
  \BibitemOpen
  \bibfield  {author} {\bibinfo {author} {\bibfnamefont {H.}~\bibnamefont {Pan}}\ and\ \bibinfo {author} {\bibfnamefont {S.~D.}\ \bibnamefont {Sarma}},\ }\href@noop {} {\bibfield  {journal} {\bibinfo  {journal} {Physical Review Research}\ }\textbf {\bibinfo {volume} {2}},\ \bibinfo {pages} {013377} (\bibinfo {year} {2020})}\BibitemShut {NoStop}%
\bibitem [{\citenamefont {Liu}\ \emph {et~al.}(2012)\citenamefont {Liu}, \citenamefont {Potter}, \citenamefont {Law},\ and\ \citenamefont {Lee}}]{liu2012zero}%
  \BibitemOpen
  \bibfield  {author} {\bibinfo {author} {\bibfnamefont {J.}~\bibnamefont {Liu}}, \bibinfo {author} {\bibfnamefont {A.~C.}\ \bibnamefont {Potter}}, \bibinfo {author} {\bibfnamefont {K.~T.}\ \bibnamefont {Law}},\ and\ \bibinfo {author} {\bibfnamefont {P.~A.}\ \bibnamefont {Lee}},\ }\href@noop {} {\bibfield  {journal} {\bibinfo  {journal} {Physical review letters}\ }\textbf {\bibinfo {volume} {109}},\ \bibinfo {pages} {267002} (\bibinfo {year} {2012})}\BibitemShut {NoStop}%
\bibitem [{\citenamefont {Das~Sarma}\ and\ \citenamefont {Pan}(2021)}]{das2021disorder}%
  \BibitemOpen
  \bibfield  {author} {\bibinfo {author} {\bibfnamefont {S.}~\bibnamefont {Das~Sarma}}\ and\ \bibinfo {author} {\bibfnamefont {H.}~\bibnamefont {Pan}},\ }\href@noop {} {\bibfield  {journal} {\bibinfo  {journal} {Physical Review B}\ }\textbf {\bibinfo {volume} {103}},\ \bibinfo {pages} {195158} (\bibinfo {year} {2021})}\BibitemShut {NoStop}%
\bibitem [{\citenamefont {Das}\ \emph {et~al.}(2012)\citenamefont {Das}, \citenamefont {Ronen}, \citenamefont {Most}, \citenamefont {Oreg}, \citenamefont {Heiblum},\ and\ \citenamefont {Shtrikman}}]{das2012zero}%
  \BibitemOpen
  \bibfield  {author} {\bibinfo {author} {\bibfnamefont {A.}~\bibnamefont {Das}}, \bibinfo {author} {\bibfnamefont {Y.}~\bibnamefont {Ronen}}, \bibinfo {author} {\bibfnamefont {Y.}~\bibnamefont {Most}}, \bibinfo {author} {\bibfnamefont {Y.}~\bibnamefont {Oreg}}, \bibinfo {author} {\bibfnamefont {M.}~\bibnamefont {Heiblum}},\ and\ \bibinfo {author} {\bibfnamefont {H.}~\bibnamefont {Shtrikman}},\ }\href@noop {} {\bibfield  {journal} {\bibinfo  {journal} {Nature Physics}\ }\textbf {\bibinfo {volume} {8}},\ \bibinfo {pages} {887} (\bibinfo {year} {2012})}\BibitemShut {NoStop}%
\bibitem [{\citenamefont {Deng}\ \emph {et~al.}(2012)\citenamefont {Deng}, \citenamefont {Yu}, \citenamefont {Huang}, \citenamefont {Larsson}, \citenamefont {Caroff},\ and\ \citenamefont {Xu}}]{deng2012anomalous}%
  \BibitemOpen
  \bibfield  {author} {\bibinfo {author} {\bibfnamefont {M.}~\bibnamefont {Deng}}, \bibinfo {author} {\bibfnamefont {C.}~\bibnamefont {Yu}}, \bibinfo {author} {\bibfnamefont {G.}~\bibnamefont {Huang}}, \bibinfo {author} {\bibfnamefont {M.}~\bibnamefont {Larsson}}, \bibinfo {author} {\bibfnamefont {P.}~\bibnamefont {Caroff}},\ and\ \bibinfo {author} {\bibfnamefont {H.}~\bibnamefont {Xu}},\ }\href@noop {} {\bibfield  {journal} {\bibinfo  {journal} {Nano letters}\ }\textbf {\bibinfo {volume} {12}},\ \bibinfo {pages} {6414} (\bibinfo {year} {2012})}\BibitemShut {NoStop}%
\bibitem [{\citenamefont {Mourik}\ \emph {et~al.}(2012)\citenamefont {Mourik}, \citenamefont {Zuo}, \citenamefont {Frolov}, \citenamefont {Plissard}, \citenamefont {Bakkers},\ and\ \citenamefont {Kouwenhoven}}]{mourik2012signatures}%
  \BibitemOpen
  \bibfield  {author} {\bibinfo {author} {\bibfnamefont {V.}~\bibnamefont {Mourik}}, \bibinfo {author} {\bibfnamefont {K.}~\bibnamefont {Zuo}}, \bibinfo {author} {\bibfnamefont {S.~M.}\ \bibnamefont {Frolov}}, \bibinfo {author} {\bibfnamefont {S.}~\bibnamefont {Plissard}}, \bibinfo {author} {\bibfnamefont {E.~P.}\ \bibnamefont {Bakkers}},\ and\ \bibinfo {author} {\bibfnamefont {L.~P.}\ \bibnamefont {Kouwenhoven}},\ }\href@noop {} {\bibfield  {journal} {\bibinfo  {journal} {Science}\ }\textbf {\bibinfo {volume} {336}},\ \bibinfo {pages} {1003} (\bibinfo {year} {2012})}\BibitemShut {NoStop}%
\bibitem [{\citenamefont {Churchill}\ \emph {et~al.}(2013)\citenamefont {Churchill}, \citenamefont {Fatemi}, \citenamefont {Grove-Rasmussen}, \citenamefont {Deng}, \citenamefont {Caroff}, \citenamefont {Xu},\ and\ \citenamefont {Marcus}}]{churchill2013superconductor}%
  \BibitemOpen
  \bibfield  {author} {\bibinfo {author} {\bibfnamefont {H.}~\bibnamefont {Churchill}}, \bibinfo {author} {\bibfnamefont {V.}~\bibnamefont {Fatemi}}, \bibinfo {author} {\bibfnamefont {K.}~\bibnamefont {Grove-Rasmussen}}, \bibinfo {author} {\bibfnamefont {M.}~\bibnamefont {Deng}}, \bibinfo {author} {\bibfnamefont {P.}~\bibnamefont {Caroff}}, \bibinfo {author} {\bibfnamefont {H.}~\bibnamefont {Xu}},\ and\ \bibinfo {author} {\bibfnamefont {C.~M.}\ \bibnamefont {Marcus}},\ }\href@noop {} {\bibfield  {journal} {\bibinfo  {journal} {Physical Review B—Condensed Matter and Materials Physics}\ }\textbf {\bibinfo {volume} {87}},\ \bibinfo {pages} {241401} (\bibinfo {year} {2013})}\BibitemShut {NoStop}%
\bibitem [{\citenamefont {Finck}\ \emph {et~al.}(2013)\citenamefont {Finck}, \citenamefont {Van~Harlingen}, \citenamefont {Mohseni}, \citenamefont {Jung},\ and\ \citenamefont {Li}}]{finck2013anomalous}%
  \BibitemOpen
  \bibfield  {author} {\bibinfo {author} {\bibfnamefont {A.}~\bibnamefont {Finck}}, \bibinfo {author} {\bibfnamefont {D.}~\bibnamefont {Van~Harlingen}}, \bibinfo {author} {\bibfnamefont {P.}~\bibnamefont {Mohseni}}, \bibinfo {author} {\bibfnamefont {K.}~\bibnamefont {Jung}},\ and\ \bibinfo {author} {\bibfnamefont {X.}~\bibnamefont {Li}},\ }\href@noop {} {\bibfield  {journal} {\bibinfo  {journal} {Physical review letters}\ }\textbf {\bibinfo {volume} {110}},\ \bibinfo {pages} {126406} (\bibinfo {year} {2013})}\BibitemShut {NoStop}%
\bibitem [{\citenamefont {Nichele}\ \emph {et~al.}(2017)\citenamefont {Nichele}, \citenamefont {Drachmann}, \citenamefont {Whiticar}, \citenamefont {O’Farrell}, \citenamefont {Suominen}, \citenamefont {Fornieri}, \citenamefont {Wang}, \citenamefont {Gardner}, \citenamefont {Thomas}, \citenamefont {Hatke} \emph {et~al.}}]{nichele2017scaling}%
  \BibitemOpen
  \bibfield  {author} {\bibinfo {author} {\bibfnamefont {F.}~\bibnamefont {Nichele}}, \bibinfo {author} {\bibfnamefont {A.~C.}\ \bibnamefont {Drachmann}}, \bibinfo {author} {\bibfnamefont {A.~M.}\ \bibnamefont {Whiticar}}, \bibinfo {author} {\bibfnamefont {E.~C.}\ \bibnamefont {O’Farrell}}, \bibinfo {author} {\bibfnamefont {H.~J.}\ \bibnamefont {Suominen}}, \bibinfo {author} {\bibfnamefont {A.}~\bibnamefont {Fornieri}}, \bibinfo {author} {\bibfnamefont {T.}~\bibnamefont {Wang}}, \bibinfo {author} {\bibfnamefont {G.~C.}\ \bibnamefont {Gardner}}, \bibinfo {author} {\bibfnamefont {C.}~\bibnamefont {Thomas}}, \bibinfo {author} {\bibfnamefont {A.~T.}\ \bibnamefont {Hatke}}, \emph {et~al.},\ }\href@noop {} {\bibfield  {journal} {\bibinfo  {journal} {Physical review letters}\ }\textbf {\bibinfo {volume} {119}},\ \bibinfo {pages} {136803} (\bibinfo {year} {2017})}\BibitemShut {NoStop}%
\bibitem [{\citenamefont {Zhang}\ \emph {et~al.}(2021)\citenamefont {Zhang}, \citenamefont {de~Moor}, \citenamefont {Bommer}, \citenamefont {Xu}, \citenamefont {Wang}, \citenamefont {van Loo}, \citenamefont {Liu}, \citenamefont {Gazibegovic}, \citenamefont {Logan}, \citenamefont {Car} \emph {et~al.}}]{zhang2021large}%
  \BibitemOpen
  \bibfield  {author} {\bibinfo {author} {\bibfnamefont {H.}~\bibnamefont {Zhang}}, \bibinfo {author} {\bibfnamefont {M.~W.}\ \bibnamefont {de~Moor}}, \bibinfo {author} {\bibfnamefont {J.~D.}\ \bibnamefont {Bommer}}, \bibinfo {author} {\bibfnamefont {D.}~\bibnamefont {Xu}}, \bibinfo {author} {\bibfnamefont {G.}~\bibnamefont {Wang}}, \bibinfo {author} {\bibfnamefont {N.}~\bibnamefont {van Loo}}, \bibinfo {author} {\bibfnamefont {C.-X.}\ \bibnamefont {Liu}}, \bibinfo {author} {\bibfnamefont {S.}~\bibnamefont {Gazibegovic}}, \bibinfo {author} {\bibfnamefont {J.~A.}\ \bibnamefont {Logan}}, \bibinfo {author} {\bibfnamefont {D.}~\bibnamefont {Car}}, \emph {et~al.},\ }\href@noop {} {\bibfield  {journal} {\bibinfo  {journal} {arXiv preprint arXiv:2101.11456}\ } (\bibinfo {year} {2021})}\BibitemShut {NoStop}%
\bibitem [{\citenamefont {Yu}\ \emph {et~al.}(2021)\citenamefont {Yu}, \citenamefont {Chen}, \citenamefont {Gomanko}, \citenamefont {Badawy}, \citenamefont {Bakkers}, \citenamefont {Zuo}, \citenamefont {Mourik},\ and\ \citenamefont {Frolov}}]{yu2021non}%
  \BibitemOpen
  \bibfield  {author} {\bibinfo {author} {\bibfnamefont {P.}~\bibnamefont {Yu}}, \bibinfo {author} {\bibfnamefont {J.}~\bibnamefont {Chen}}, \bibinfo {author} {\bibfnamefont {M.}~\bibnamefont {Gomanko}}, \bibinfo {author} {\bibfnamefont {G.}~\bibnamefont {Badawy}}, \bibinfo {author} {\bibfnamefont {E.}~\bibnamefont {Bakkers}}, \bibinfo {author} {\bibfnamefont {K.}~\bibnamefont {Zuo}}, \bibinfo {author} {\bibfnamefont {V.}~\bibnamefont {Mourik}},\ and\ \bibinfo {author} {\bibfnamefont {S.}~\bibnamefont {Frolov}},\ }\href@noop {} {\bibfield  {journal} {\bibinfo  {journal} {Nature Physics}\ }\textbf {\bibinfo {volume} {17}},\ \bibinfo {pages} {482} (\bibinfo {year} {2021})}\BibitemShut {NoStop}%
\bibitem [{\citenamefont {Song}\ \emph {et~al.}(2022)\citenamefont {Song}, \citenamefont {Zhang}, \citenamefont {Pan}, \citenamefont {Liu}, \citenamefont {Wang}, \citenamefont {Cao}, \citenamefont {Liu}, \citenamefont {Wen}, \citenamefont {Liao}, \citenamefont {Zhuo} \emph {et~al.}}]{song2022large}%
  \BibitemOpen
  \bibfield  {author} {\bibinfo {author} {\bibfnamefont {H.}~\bibnamefont {Song}}, \bibinfo {author} {\bibfnamefont {Z.}~\bibnamefont {Zhang}}, \bibinfo {author} {\bibfnamefont {D.}~\bibnamefont {Pan}}, \bibinfo {author} {\bibfnamefont {D.}~\bibnamefont {Liu}}, \bibinfo {author} {\bibfnamefont {Z.}~\bibnamefont {Wang}}, \bibinfo {author} {\bibfnamefont {Z.}~\bibnamefont {Cao}}, \bibinfo {author} {\bibfnamefont {L.}~\bibnamefont {Liu}}, \bibinfo {author} {\bibfnamefont {L.}~\bibnamefont {Wen}}, \bibinfo {author} {\bibfnamefont {D.}~\bibnamefont {Liao}}, \bibinfo {author} {\bibfnamefont {R.}~\bibnamefont {Zhuo}}, \emph {et~al.},\ }\href@noop {} {\bibfield  {journal} {\bibinfo  {journal} {Physical Review Research}\ }\textbf {\bibinfo {volume} {4}},\ \bibinfo {pages} {033235} (\bibinfo {year} {2022})}\BibitemShut {NoStop}%
\bibitem [{\citenamefont {Aghaee}\ \emph {et~al.}(2023)\citenamefont {Aghaee}, \citenamefont {Akkala}, \citenamefont {Alam}, \citenamefont {Ali}, \citenamefont {Ramirez}, \citenamefont {Andrzejczuk}, \citenamefont {Antipov}, \citenamefont {Aseev}, \citenamefont {Astafev}, \citenamefont {Bauer} \emph {et~al.}}]{aghaee2023inas}%
  \BibitemOpen
  \bibfield  {author} {\bibinfo {author} {\bibfnamefont {M.}~\bibnamefont {Aghaee}}, \bibinfo {author} {\bibfnamefont {A.}~\bibnamefont {Akkala}}, \bibinfo {author} {\bibfnamefont {Z.}~\bibnamefont {Alam}}, \bibinfo {author} {\bibfnamefont {R.}~\bibnamefont {Ali}}, \bibinfo {author} {\bibfnamefont {A.~A.}\ \bibnamefont {Ramirez}}, \bibinfo {author} {\bibfnamefont {M.}~\bibnamefont {Andrzejczuk}}, \bibinfo {author} {\bibfnamefont {A.~E.}\ \bibnamefont {Antipov}}, \bibinfo {author} {\bibfnamefont {P.}~\bibnamefont {Aseev}}, \bibinfo {author} {\bibfnamefont {M.}~\bibnamefont {Astafev}}, \bibinfo {author} {\bibfnamefont {B.}~\bibnamefont {Bauer}}, \emph {et~al.},\ }\href@noop {} {\bibfield  {journal} {\bibinfo  {journal} {Physical Review B}\ }\textbf {\bibinfo {volume} {107}},\ \bibinfo {pages} {245423} (\bibinfo {year} {2023})}\BibitemShut {NoStop}%
\bibitem [{\citenamefont {Pikulin}\ \emph {et~al.}(2021)\citenamefont {Pikulin}, \citenamefont {van Heck}, \citenamefont {Karzig}, \citenamefont {Martinez}, \citenamefont {Nijholt}, \citenamefont {Laeven}, \citenamefont {Winkler}, \citenamefont {Watson}, \citenamefont {Heedt}, \citenamefont {Temurhan} \emph {et~al.}}]{pikulin2021protocol}%
  \BibitemOpen
  \bibfield  {author} {\bibinfo {author} {\bibfnamefont {D.~I.}\ \bibnamefont {Pikulin}}, \bibinfo {author} {\bibfnamefont {B.}~\bibnamefont {van Heck}}, \bibinfo {author} {\bibfnamefont {T.}~\bibnamefont {Karzig}}, \bibinfo {author} {\bibfnamefont {E.~A.}\ \bibnamefont {Martinez}}, \bibinfo {author} {\bibfnamefont {B.}~\bibnamefont {Nijholt}}, \bibinfo {author} {\bibfnamefont {T.}~\bibnamefont {Laeven}}, \bibinfo {author} {\bibfnamefont {G.~W.}\ \bibnamefont {Winkler}}, \bibinfo {author} {\bibfnamefont {J.~D.}\ \bibnamefont {Watson}}, \bibinfo {author} {\bibfnamefont {S.}~\bibnamefont {Heedt}}, \bibinfo {author} {\bibfnamefont {M.}~\bibnamefont {Temurhan}}, \emph {et~al.},\ }\href@noop {} {\bibfield  {journal} {\bibinfo  {journal} {arXiv preprint arXiv:2103.12217}\ } (\bibinfo {year} {2021})}\BibitemShut {NoStop}%
\bibitem [{\citenamefont {Pan}\ \emph {et~al.}(2021)\citenamefont {Pan}, \citenamefont {Sau},\ and\ \citenamefont {Das~Sarma}}]{pan2021three}%
  \BibitemOpen
  \bibfield  {author} {\bibinfo {author} {\bibfnamefont {H.}~\bibnamefont {Pan}}, \bibinfo {author} {\bibfnamefont {J.~D.}\ \bibnamefont {Sau}},\ and\ \bibinfo {author} {\bibfnamefont {S.}~\bibnamefont {Das~Sarma}},\ }\href@noop {} {\bibfield  {journal} {\bibinfo  {journal} {Physical Review B}\ }\textbf {\bibinfo {volume} {103}},\ \bibinfo {pages} {014513} (\bibinfo {year} {2021})}\BibitemShut {NoStop}%
\bibitem [{\citenamefont {Das~Sarma}\ \emph {et~al.}(2023)\citenamefont {Das~Sarma}, \citenamefont {Sau},\ and\ \citenamefont {Stanescu}}]{das2023spectral}%
  \BibitemOpen
  \bibfield  {author} {\bibinfo {author} {\bibfnamefont {S.}~\bibnamefont {Das~Sarma}}, \bibinfo {author} {\bibfnamefont {J.~D.}\ \bibnamefont {Sau}},\ and\ \bibinfo {author} {\bibfnamefont {T.~D.}\ \bibnamefont {Stanescu}},\ }\href@noop {} {\bibfield  {journal} {\bibinfo  {journal} {Physical Review B}\ }\textbf {\bibinfo {volume} {108}},\ \bibinfo {pages} {085416} (\bibinfo {year} {2023})}\BibitemShut {NoStop}%
\bibitem [{\citenamefont {Das~Sarma}\ and\ \citenamefont {Pan}(2023)}]{das2023density}%
  \BibitemOpen
  \bibfield  {author} {\bibinfo {author} {\bibfnamefont {S.}~\bibnamefont {Das~Sarma}}\ and\ \bibinfo {author} {\bibfnamefont {H.}~\bibnamefont {Pan}},\ }\href@noop {} {\bibfield  {journal} {\bibinfo  {journal} {Physical Review B}\ }\textbf {\bibinfo {volume} {108}},\ \bibinfo {pages} {085415} (\bibinfo {year} {2023})}\BibitemShut {NoStop}%
\bibitem [{\citenamefont {Akhmerov}\ \emph {et~al.}(2011)\citenamefont {Akhmerov}, \citenamefont {Dahlhaus}, \citenamefont {Hassler}, \citenamefont {Wimmer},\ and\ \citenamefont {Beenakker}}]{akhmerov2011quantized}%
  \BibitemOpen
  \bibfield  {author} {\bibinfo {author} {\bibfnamefont {A.}~\bibnamefont {Akhmerov}}, \bibinfo {author} {\bibfnamefont {J.}~\bibnamefont {Dahlhaus}}, \bibinfo {author} {\bibfnamefont {F.}~\bibnamefont {Hassler}}, \bibinfo {author} {\bibfnamefont {M.}~\bibnamefont {Wimmer}},\ and\ \bibinfo {author} {\bibfnamefont {C.}~\bibnamefont {Beenakker}},\ }\href@noop {} {\bibfield  {journal} {\bibinfo  {journal} {Physical review letters}\ }\textbf {\bibinfo {volume} {106}},\ \bibinfo {pages} {057001} (\bibinfo {year} {2011})}\BibitemShut {NoStop}%
\bibitem [{\citenamefont {Pan}\ and\ \citenamefont {Das~Sarma}(2024)}]{pan2024disordered}%
  \BibitemOpen
  \bibfield  {author} {\bibinfo {author} {\bibfnamefont {H.}~\bibnamefont {Pan}}\ and\ \bibinfo {author} {\bibfnamefont {S.}~\bibnamefont {Das~Sarma}},\ }\href@noop {} {\bibfield  {journal} {\bibinfo  {journal} {Physical Review B}\ }\textbf {\bibinfo {volume} {110}},\ \bibinfo {pages} {075401} (\bibinfo {year} {2024})}\BibitemShut {NoStop}%
\bibitem [{\citenamefont {Taylor}\ and\ \citenamefont {Das~Sarma}(2025{\natexlab{a}})}]{taylor2025vision}%
  \BibitemOpen
  \bibfield  {author} {\bibinfo {author} {\bibfnamefont {J.~R.}\ \bibnamefont {Taylor}}\ and\ \bibinfo {author} {\bibfnamefont {S.}~\bibnamefont {Das~Sarma}},\ }\href@noop {} {\bibfield  {journal} {\bibinfo  {journal} {Physical Review B}\ }\textbf {\bibinfo {volume} {111}},\ \bibinfo {pages} {104208} (\bibinfo {year} {2025}{\natexlab{a}})}\BibitemShut {NoStop}%
\bibitem [{\citenamefont {Taylor}\ \emph {et~al.}(2024)\citenamefont {Taylor}, \citenamefont {Sau},\ and\ \citenamefont {Das~Sarma}}]{taylor2024machine}%
  \BibitemOpen
  \bibfield  {author} {\bibinfo {author} {\bibfnamefont {J.~R.}\ \bibnamefont {Taylor}}, \bibinfo {author} {\bibfnamefont {J.~D.}\ \bibnamefont {Sau}},\ and\ \bibinfo {author} {\bibfnamefont {S.}~\bibnamefont {Das~Sarma}},\ }\href@noop {} {\bibfield  {journal} {\bibinfo  {journal} {Physical Review Letters}\ }\textbf {\bibinfo {volume} {132}},\ \bibinfo {pages} {206602} (\bibinfo {year} {2024})}\BibitemShut {NoStop}%
\bibitem [{\citenamefont {Cheng}\ \emph {et~al.}(2024)\citenamefont {Cheng}, \citenamefont {Okabe}, \citenamefont {Chotrattanapituk},\ and\ \citenamefont {Li}}]{cheng2024machine}%
  \BibitemOpen
  \bibfield  {author} {\bibinfo {author} {\bibfnamefont {M.}~\bibnamefont {Cheng}}, \bibinfo {author} {\bibfnamefont {R.}~\bibnamefont {Okabe}}, \bibinfo {author} {\bibfnamefont {A.}~\bibnamefont {Chotrattanapituk}},\ and\ \bibinfo {author} {\bibfnamefont {M.}~\bibnamefont {Li}},\ }\href@noop {} {\bibfield  {journal} {\bibinfo  {journal} {Matter}\ }\textbf {\bibinfo {volume} {7}},\ \bibinfo {pages} {2507 } (\bibinfo {year} {2024})}\BibitemShut {NoStop}%
\bibitem [{\citenamefont {Thamm}\ and\ \citenamefont {Rosenow}(2023)}]{thamm2023machine}%
  \BibitemOpen
  \bibfield  {author} {\bibinfo {author} {\bibfnamefont {M.}~\bibnamefont {Thamm}}\ and\ \bibinfo {author} {\bibfnamefont {B.}~\bibnamefont {Rosenow}},\ }\href@noop {} {\bibfield  {journal} {\bibinfo  {journal} {Physical Review Letters}\ }\textbf {\bibinfo {volume} {130}},\ \bibinfo {pages} {116202} (\bibinfo {year} {2023})}\BibitemShut {NoStop}%
\bibitem [{\citenamefont {Thamm}\ and\ \citenamefont {Rosenow}(2024)}]{thamm2024conductance}%
  \BibitemOpen
  \bibfield  {author} {\bibinfo {author} {\bibfnamefont {M.}~\bibnamefont {Thamm}}\ and\ \bibinfo {author} {\bibfnamefont {B.}~\bibnamefont {Rosenow}},\ }\href@noop {} {\bibfield  {journal} {\bibinfo  {journal} {Physical Review B}\ }\textbf {\bibinfo {volume} {109}},\ \bibinfo {pages} {045132} (\bibinfo {year} {2024})}\BibitemShut {NoStop}%
\bibitem [{\citenamefont {Hansen}(2016)}]{hansen2016cma}%
  \BibitemOpen
  \bibfield  {author} {\bibinfo {author} {\bibfnamefont {N.}~\bibnamefont {Hansen}},\ }\href@noop {} {\bibfield  {journal} {\bibinfo  {journal} {arXiv preprint arXiv:1604.00772}\ } (\bibinfo {year} {2016})}\BibitemShut {NoStop}%
\bibitem [{\citenamefont {Dosovitskiy}\ \emph {et~al.}(2020)\citenamefont {Dosovitskiy}, \citenamefont {Beyer}, \citenamefont {Kolesnikov}, \citenamefont {Weissenborn}, \citenamefont {Zhai}, \citenamefont {Unterthiner}, \citenamefont {Dehghani}, \citenamefont {Minderer}, \citenamefont {Heigold}, \citenamefont {Gelly} \emph {et~al.}}]{dosovitskiy2020image}%
  \BibitemOpen
  \bibfield  {author} {\bibinfo {author} {\bibfnamefont {A.}~\bibnamefont {Dosovitskiy}}, \bibinfo {author} {\bibfnamefont {L.}~\bibnamefont {Beyer}}, \bibinfo {author} {\bibfnamefont {A.}~\bibnamefont {Kolesnikov}}, \bibinfo {author} {\bibfnamefont {D.}~\bibnamefont {Weissenborn}}, \bibinfo {author} {\bibfnamefont {X.}~\bibnamefont {Zhai}}, \bibinfo {author} {\bibfnamefont {T.}~\bibnamefont {Unterthiner}}, \bibinfo {author} {\bibfnamefont {M.}~\bibnamefont {Dehghani}}, \bibinfo {author} {\bibfnamefont {M.}~\bibnamefont {Minderer}}, \bibinfo {author} {\bibfnamefont {G.}~\bibnamefont {Heigold}}, \bibinfo {author} {\bibfnamefont {S.}~\bibnamefont {Gelly}}, \emph {et~al.},\ }\href@noop {} {\bibfield  {journal} {\bibinfo  {journal} {arXiv preprint arXiv:2010.11929}\ } (\bibinfo {year} {2020})}\BibitemShut {NoStop}%
\bibitem [{\citenamefont {Taylor}\ and\ \citenamefont {Das~Sarma}(2025{\natexlab{b}})}]{taylor2024neural}%
  \BibitemOpen
  \bibfield  {author} {\bibinfo {author} {\bibfnamefont {J.~R.}\ \bibnamefont {Taylor}}\ and\ \bibinfo {author} {\bibfnamefont {S.}~\bibnamefont {Das~Sarma}},\ }\href {https://doi.org/10.1103/PhysRevB.111.035301} {\bibfield  {journal} {\bibinfo  {journal} {Phys. Rev. B}\ }\textbf {\bibinfo {volume} {111}},\ \bibinfo {pages} {035301} (\bibinfo {year} {2025}{\natexlab{b}})}\BibitemShut {NoStop}%
\bibitem [{\citenamefont {Wolz}(2021)}]{fcmaes2021}%
  \BibitemOpen
  \bibfield  {author} {\bibinfo {author} {\bibfnamefont {D.}~\bibnamefont {Wolz}},\ }\href@noop {} {\bibinfo {title} {fcmaes - a python 3 gradient-free optimization library}},\ \bibinfo {howpublished} {\url{https://github.com/dietmarwo/fast-cma-es}} (\bibinfo {year} {2021}),\ \bibinfo {note} {python/C++ source code}\BibitemShut {NoStop}%
\bibitem [{\citenamefont {Groth}\ \emph {et~al.}(2014)\citenamefont {Groth}, \citenamefont {Wimmer}, \citenamefont {Akhmerov},\ and\ \citenamefont {Waintal}}]{groth2014kwant}%
  \BibitemOpen
  \bibfield  {author} {\bibinfo {author} {\bibfnamefont {C.~W.}\ \bibnamefont {Groth}}, \bibinfo {author} {\bibfnamefont {M.}~\bibnamefont {Wimmer}}, \bibinfo {author} {\bibfnamefont {A.~R.}\ \bibnamefont {Akhmerov}},\ and\ \bibinfo {author} {\bibfnamefont {X.}~\bibnamefont {Waintal}},\ }\href@noop {} {\bibfield  {journal} {\bibinfo  {journal} {New Journal of Physics}\ }\textbf {\bibinfo {volume} {16}},\ \bibinfo {pages} {063065} (\bibinfo {year} {2014})}\BibitemShut {NoStop}%
\end{thebibliography}%


\begin{thebibliography}{7}%
\makeatletter
\providecommand \@ifxundefined [1]{%
 \@ifx{#1\undefined}
}%
\providecommand \@ifnum [1]{%
 \ifnum #1\expandafter \@firstoftwo
 \else \expandafter \@secondoftwo
 \fi
}%
\providecommand \@ifx [1]{%
 \ifx #1\expandafter \@firstoftwo
 \else \expandafter \@secondoftwo
 \fi
}%
\providecommand \natexlab [1]{#1}%
\providecommand \enquote  [1]{``#1''}%
\providecommand \bibnamefont  [1]{#1}%
\providecommand \bibfnamefont [1]{#1}%
\providecommand \citenamefont [1]{#1}%
\providecommand \href@noop [0]{\@secondoftwo}%
\providecommand \href [0]{\begingroup \@sanitize@url \@href}%
\providecommand \@href[1]{\@@startlink{#1}\@@href}%
\providecommand \@@href[1]{\endgroup#1\@@endlink}%
\providecommand \@sanitize@url [0]{\catcode `\\12\catcode `\$12\catcode `\&12\catcode `\#12\catcode `\^12\catcode `\_12\catcode `\%12\relax}%
\providecommand \@@startlink[1]{}%
\providecommand \@@endlink[0]{}%
\providecommand \url  [0]{\begingroup\@sanitize@url \@url }%
\providecommand \@url [1]{\endgroup\@href {#1}{\urlprefix }}%
\providecommand \urlprefix  [0]{URL }%
\providecommand \Eprint [0]{\href }%
\providecommand \doibase [0]{https://doi.org/}%
\providecommand \selectlanguage [0]{\@gobble}%
\providecommand \bibinfo  [0]{\@secondoftwo}%
\providecommand \bibfield  [0]{\@secondoftwo}%
\providecommand \translation [1]{[#1]}%
\providecommand \BibitemOpen [0]{}%
\providecommand \bibitemStop [0]{}%
\providecommand \bibitemNoStop [0]{.\EOS\space}%
\providecommand \EOS [0]{\spacefactor3000\relax}%
\providecommand \BibitemShut  [1]{\csname bibitem#1\endcsname}%
\let\auto@bib@innerbib\@empty
\bibitem [{\citenamefont {Taylor}\ and\ \citenamefont {Das~Sarma}(2025{\natexlab{a}})}]{taylor2025vision}%
  \BibitemOpen
  \bibfield  {author} {\bibinfo {author} {\bibfnamefont {J.~R.}\ \bibnamefont {Taylor}}\ and\ \bibinfo {author} {\bibfnamefont {S.}~\bibnamefont {Das~Sarma}},\ }\href@noop {} {\bibfield  {journal} {\bibinfo  {journal} {Physical Review B}\ }\textbf {\bibinfo {volume} {111}},\ \bibinfo {pages} {104208} (\bibinfo {year} {2025}{\natexlab{a}})}\BibitemShut {NoStop}%
\bibitem [{\citenamefont {Dosovitskiy}\ \emph {et~al.}(2020)\citenamefont {Dosovitskiy}, \citenamefont {Beyer}, \citenamefont {Kolesnikov}, \citenamefont {Weissenborn}, \citenamefont {Zhai}, \citenamefont {Unterthiner}, \citenamefont {Dehghani}, \citenamefont {Minderer}, \citenamefont {Heigold}, \citenamefont {Gelly} \emph {et~al.}}]{dosovitskiy2020image}%
  \BibitemOpen
  \bibfield  {author} {\bibinfo {author} {\bibfnamefont {A.}~\bibnamefont {Dosovitskiy}}, \bibinfo {author} {\bibfnamefont {L.}~\bibnamefont {Beyer}}, \bibinfo {author} {\bibfnamefont {A.}~\bibnamefont {Kolesnikov}}, \bibinfo {author} {\bibfnamefont {D.}~\bibnamefont {Weissenborn}}, \bibinfo {author} {\bibfnamefont {X.}~\bibnamefont {Zhai}}, \bibinfo {author} {\bibfnamefont {T.}~\bibnamefont {Unterthiner}}, \bibinfo {author} {\bibfnamefont {M.}~\bibnamefont {Dehghani}}, \bibinfo {author} {\bibfnamefont {M.}~\bibnamefont {Minderer}}, \bibinfo {author} {\bibfnamefont {G.}~\bibnamefont {Heigold}}, \bibinfo {author} {\bibfnamefont {S.}~\bibnamefont {Gelly}}, \emph {et~al.},\ }\href@noop {} {\bibfield  {journal} {\bibinfo  {journal} {arXiv preprint arXiv:2010.11929}\ } (\bibinfo {year} {2020})}\BibitemShut {NoStop}%
\bibitem [{\citenamefont {Taylor}\ \emph {et~al.}(2024)\citenamefont {Taylor}, \citenamefont {Sau},\ and\ \citenamefont {Das~Sarma}}]{taylor2024machine}%
  \BibitemOpen
  \bibfield  {author} {\bibinfo {author} {\bibfnamefont {J.~R.}\ \bibnamefont {Taylor}}, \bibinfo {author} {\bibfnamefont {J.~D.}\ \bibnamefont {Sau}},\ and\ \bibinfo {author} {\bibfnamefont {S.}~\bibnamefont {Das~Sarma}},\ }\href@noop {} {\bibfield  {journal} {\bibinfo  {journal} {Physical Review Letters}\ }\textbf {\bibinfo {volume} {132}},\ \bibinfo {pages} {206602} (\bibinfo {year} {2024})}\BibitemShut {NoStop}%
\bibitem [{\citenamefont {Taylor}\ and\ \citenamefont {Das~Sarma}(2025{\natexlab{b}})}]{taylor2024neural}%
  \BibitemOpen
  \bibfield  {author} {\bibinfo {author} {\bibfnamefont {J.~R.}\ \bibnamefont {Taylor}}\ and\ \bibinfo {author} {\bibfnamefont {S.}~\bibnamefont {Das~Sarma}},\ }\href {https://doi.org/10.1103/PhysRevB.111.035301} {\bibfield  {journal} {\bibinfo  {journal} {Phys. Rev. B}\ }\textbf {\bibinfo {volume} {111}},\ \bibinfo {pages} {035301} (\bibinfo {year} {2025}{\natexlab{b}})}\BibitemShut {NoStop}%
\bibitem [{\citenamefont {Groth}\ \emph {et~al.}(2014)\citenamefont {Groth}, \citenamefont {Wimmer}, \citenamefont {Akhmerov},\ and\ \citenamefont {Waintal}}]{groth2014kwant}%
  \BibitemOpen
  \bibfield  {author} {\bibinfo {author} {\bibfnamefont {C.~W.}\ \bibnamefont {Groth}}, \bibinfo {author} {\bibfnamefont {M.}~\bibnamefont {Wimmer}}, \bibinfo {author} {\bibfnamefont {A.~R.}\ \bibnamefont {Akhmerov}},\ and\ \bibinfo {author} {\bibfnamefont {X.}~\bibnamefont {Waintal}},\ }\href@noop {} {\bibfield  {journal} {\bibinfo  {journal} {New Journal of Physics}\ }\textbf {\bibinfo {volume} {16}},\ \bibinfo {pages} {063065} (\bibinfo {year} {2014})}\BibitemShut {NoStop}%
\bibitem [{\citenamefont {Hansen}(2016)}]{hansen2016cma}%
  \BibitemOpen
  \bibfield  {author} {\bibinfo {author} {\bibfnamefont {N.}~\bibnamefont {Hansen}},\ }\href@noop {} {\bibfield  {journal} {\bibinfo  {journal} {arXiv preprint arXiv:1604.00772}\ } (\bibinfo {year} {2016})}\BibitemShut {NoStop}%
\bibitem [{\citenamefont {Wolz}(2021)}]{fcmaes2021}%
  \BibitemOpen
  \bibfield  {author} {\bibinfo {author} {\bibfnamefont {D.}~\bibnamefont {Wolz}},\ }\href@noop {} {\bibinfo {title} {fcmaes - a python 3 gradient-free optimization library}},\ \bibinfo {howpublished} {\url{https://github.com/dietmarwo/fast-cma-es}} (\bibinfo {year} {2021}),\ \bibinfo {note} {python/C++ source code}\BibitemShut {NoStop}%
\end{thebibliography}%


\begin{thebibliography}{0}%
\makeatletter
\providecommand \@ifxundefined [1]{%
 \@ifx{#1\undefined}
}%
\providecommand \@ifnum [1]{%
 \ifnum #1\expandafter \@firstoftwo
 \else \expandafter \@secondoftwo
 \fi
}%
\providecommand \@ifx [1]{%
 \ifx #1\expandafter \@firstoftwo
 \else \expandafter \@secondoftwo
 \fi
}%
\providecommand \natexlab [1]{#1}%
\providecommand \enquote  [1]{``#1''}%
\providecommand \bibnamefont  [1]{#1}%
\providecommand \bibfnamefont [1]{#1}%
\providecommand \citenamefont [1]{#1}%
\providecommand \href@noop [0]{\@secondoftwo}%
\providecommand \href [0]{\begingroup \@sanitize@url \@href}%
\providecommand \@href[1]{\@@startlink{#1}\@@href}%
\providecommand \@@href[1]{\endgroup#1\@@endlink}%
\providecommand \@sanitize@url [0]{\catcode `\\12\catcode `\$12\catcode `\&12\catcode `\#12\catcode `\^12\catcode `\_12\catcode `\%12\relax}%
\providecommand \@@startlink[1]{}%
\providecommand \@@endlink[0]{}%
\providecommand \url  [0]{\begingroup\@sanitize@url \@url }%
\providecommand \@url [1]{\endgroup\@href {#1}{\urlprefix }}%
\providecommand \urlprefix  [0]{URL }%
\providecommand \Eprint [0]{\href }%
\providecommand \doibase [0]{https://doi.org/}%
\providecommand \selectlanguage [0]{\@gobble}%
\providecommand \bibinfo  [0]{\@secondoftwo}%
\providecommand \bibfield  [0]{\@secondoftwo}%
\providecommand \translation [1]{[#1]}%
\providecommand \BibitemOpen [0]{}%
\providecommand \bibitemStop [0]{}%
\providecommand \bibitemNoStop [0]{.\EOS\space}%
\providecommand \EOS [0]{\spacefactor3000\relax}%
\providecommand \BibitemShut  [1]{\csname bibitem#1\endcsname}%
\let\auto@bib@innerbib\@empty
\end{thebibliography}%
\end{document}